\documentclass[prd,11pt, showpacs]{revtex4-1}

\usepackage{amsmath}    
\usepackage{graphicx}   
\usepackage{verbatim}   
\usepackage{color}      
\usepackage{subfigure}  
\usepackage{hyperref}   
\usepackage{afterpage,float,tensor,enumerate,mhchem,dsfont}
\usepackage{xfrac,mathtools,amssymb,amsthm,braket,multirow}
\usepackage{slashed,appendix,listings}
\newcommand{\abs}[1]{\left|#1\right|}
\begin{document}

\title{A quantum field theoretical model of neutrino oscillation without external wave packets}
\author{Z. Y. Law}
\email[Email:]{lawzhiyang@yahoo.com}
\author{A. H. Chan}
\email[Email:]{phycahp@nus.edu.sg}
\author{C. H. Oh}
\email[Email:]{phyohch@nus.edu.sg }
\affiliation{Department of Physics, National University of Singapore, 2 Science Drive, Singapore 115742, Singapore
}

\date{\today}
\begin{abstract}
   We develop a general and consistent model of neutrino oscillation based on the quantum field theoretical (QFT) description of the neutrino production and detection processes. Emphasis is placed on the locality of the interactions of these processes, where on top of the usual application of the 4-Fermion local Hamiltonian, we assume that weak interactions “switched on” only when the wave functions of the particles involved are overlapping and “switched off” upon their separation.

A key assumption in our treatment is that the wave packet sizes of the particles, in particular, the neutrino-producing source particles and the neutrino absorbing detector particles, are taken to be negligible compared with their mean free path in their respective medium. With this assumption, and taking into considerations of the finite time of neutrino production, neutrino wave packets with well defined edges are generated. This fact, together with the locality of weak interactions, enable us to relate the propagation time to the propagation distance, thus doing away with the ad hoc time averaging procedure normally employing in derivations of neutrino oscillation formula. No ansatzs on the particular forms of particle wave functions (for example, gaussians) need to be presupposed. 

As a result, for the case of ultra-relativistic neutrinos, we have derived a neutrino count rate formula, which will be useful for making direct connection to experiments. A neutrino flavor oscillation probability arises naturally from this formula, which, when compared to the standard oscillation formula, shows modifications dependent on the relative velocities of the source/detector particles and the decoherence time taken for the unstable source to collapse into its undecayed or decayed states. This correction could be significant for short baseline neutrino oscillation experiments. A good feature of our approach is that the neutrino oscillation formula is automatically normalized if the in-going states of the production and detection processes are normalized. 

We also show that causality and unitarity cannot both be satisfied in virtual neutrino models.  
\end{abstract}

\pacs{14.60.Pq, 13.15.+g, 12.15.Ff}
\maketitle
\section{Introduction}

   With the discovery of the Higgs particle, all the elements of the Standard Model are now complete, and neutrino masses remain as the sole tangible indication of physics beyond the Standard Model. This window into new physics is built on the foundation of the phenomenon of neutrino flavor oscillations, which remain to be the chief avenue through which experimental data on neutrino masses and mixing parameters are gathered. 

Given the inherent challenges involved in conducting neutrino oscillation experiments, and the equally stringent standards required in seeking experimental evidences for new physics; it is important that the steps from any theoretical mass models of the neutrino leading to the neutrino oscillation probability formula \cite{Z. Z. Xing}, which is to make direct connections with experiments, be free of ambiguities and errors. 

The main aim of this paper is to strengthen this link and provides a more rigorous derivation of the neutrino oscillation formula, while identifying the conditions under which deviations from the standard oscillation formula may be expected.

This is not the first attempt towards such a goal. From the original derivation presented by Pontecorvo \cite{Pontecorvo}, based on the planewave description of the neutrinos, modifications have been suggested to give a more physical description of the neutrino itself, by describing them with wave packets \cite{Nussinov,Kayser1981,Giunti1991,Kiers} in the quantum mechanical formalism, to mirror the way neutrinos are produced and detected locally. These wave packets are put in by hand without considerations as to how they are formed. This gives a set of conditions under which neutrino oscillation can occur, namely, there exist a coherence length and that the neutrino wave packet must have a size small compared to the oscillation length. This last condition rules out the plane wave description.

Further investigations \cite{Akhmedov,Kayser2010,Cohen,Akhmedov2009,Giunti2002,Dolgov,Beuthe,Jun Wu}, taking into considerations the production processes of the neutrinos, reveal that neutrinos are produced in weak decays in an entangled state with the associated charged lepton, and that on tracing out the lepton state (or the rest of the decay products), the remaining neutrino state can only be plane waves. This reveals the problem of disentangling the neutrino while maintaining its wave packet form.

The most common way around this problem in the quantum field theoretical treatment of neutrino oscillation is, instead of tracing out the lepton state from the joint neutrino-lepton decayed (from an unstable parent) state in some basis (for example eigen-states), the lepton is assumed to be measured as a wave packet. Thus, a corresponding wave packet is induced, through entanglement, on the neutrino state.

This is done for the external wave packet model (see \cite{Beuthe} and the references therein), in which the neutrino is described as a virtual particle (therefore, not an “external” particle; see for example \cite{Grimus1996,Grimus1999,Grimus2000}), and all external particles are each assigned a wave packet. For the intermediate wave packet model \cite{Beuthe}  (for example \cite{Giunti2002}), in which a neutrino is treated as a real particle, the neutrino is simply assumed to exist as a wave packet, as in the quantum mechanical wave packet treatment, and thus does not address the issue of neutrino localization.             

We feel that since in most of the treatments just mentioned, the neutrino oscillation formula is dependent on the sizes of the particle wave packets, including the measured lepton wave packet; this means that the ways in which the leptons are measured (for example, the spatial resolution of the detector measuring it) must be discernable by the neutrino observer. In a hypothetical neutrino-lepton EPR (Einstein-Podolsky-Rosen) scenario, causality might be violated (effects of the measurement of the recoil particle on neutrino oscillation through entanglement is also discussed in \cite{Akhmedov} and \cite{Kayser2010}, with the conclusion that neutrino oscillation is not observable if phases due to the recoil particle is significant; but this effect would be discernable to the neutrino detector).

Thus, to avoid potential complications, we build a full quantum field theoretical model based on local interactions (for example the 4 fermion weak interactions), assuming wave packets only for the source particle (producing neutrinos through decay) and the detector particle (absorbs the neutrino to give a lepton). The neutrino wave packet in our model results from the decay process itself, as the lepton state is traced out in a convenient basis (momentum basis). 

We consider a decoherence time ($T_{1}$) between the decayed and the undecayed state of the source, after which the neutrino is emitted (source collapsed to the decayed state), and propagates for the time, $T_{prop}$, when the neutrino wave function overlap with that of the detector particle and the detection interaction begins.

During this interaction, the neutrino-detector state evolves into a superposition of the detection (neutrino absorbed) and non-detection (neutrino not absorbed) state, which collapsed into the former at the end of the time interval $T_{2}$ (decoherence time associated with detection). Thus, we are concerned with 3 time scales, $T_{1}$, $T_{prop}$ and $T_{2}$ which are present in all neutrino oscillation scenario, and show how they enter into the neutrino oscillation formula.

We assume no particular forms for the wave packets describing the source and the detector particles, except that their sizes are much smaller than the decoherence time scales ($T_{1}$ and $T_{2}$). Our model is considered as an intermediate wave packet model.

More examples of earlier works done on the field theoretical treatment of neutrino can be found in \cite{Akhmedov,Akhmedov2009,Pallin,Giunti2002,Fujii,Cardall,Dolgov,Beuthe,Grimus1996,Grimus1999,Grimus2000,Jun Wu}, which include considerations of neutrino propagators both in vacuum and matter \cite{Pallin,Cardall}, as well as the possibility of constructing the Hilbert space for neutrino flavor states \cite{Pallin}.

In this paper, we begin by highlighting a potential problem which occurs in modeling neutrino oscillation as a process involving the exchange of virtual neutrinos between a source and a detector (which is common in such field theoretical models). It is shown that there is a conflict between unitary and causality in the probabilities obtain in such treatments. This will serve as a motivation for the development of a model that is free from such problems, which will be the subject of the rest of the paper.

We conclude with the derivation of a neutrino count rate formula, which will be useful for making direct connection with experiments, followed by a discussion of results.  

 \section{The problem with virtual neutrino models}       

As mentioned in the last section, virtual neutrino models consider neutrino oscillation as a single process including both the neutrino production and detection interactions. To calculate the probability for such a process, one computes the corresponding Feynman as shown in Fig. 1. $S_{in}/D_{in}$ (source/detector) is the in-going states of the scattering, that include the state of the source which will emit the neutrino, and the state of the detector before neutrino detection. $S_{out}/D_{out}$ represents the products of this interaction. The neutrino is the propagator connecting the production/detection vertices located at the space-time points, $y$/$x$, while $\mathcal{V}_{S/D}$ is the space-time uncertainties of the production/detection. We shall show that under such a context, this ampltiude cannot be both causal and unitary. 

By causal, we mean the energy resolution of the process allows us to distinguish the source (emitting energy/neutrino) from the detector (receiving energy/neutrino); this naturally implies a temporal ordering (emission before detection), since the energy of the neutrino is positive. Causality is ensured in such models by choosing the appropriate external wave packets (wave functions of the in/out-going particles) such that $y_{0}>x_{0}$ and $\mathcal{V}_{S/D}$ do not overlap.

The unitarity condition is given by
\begin{equation}
\bra{f}iT\ket{i}+\bra{f}(iT)^{\dagger}\ket{i}=-\sum_{n}\bra{f}(iT)^{\dagger}\ket{n}\bra{n}(iT)\ket{i}
\end{equation}    
where $T$ is the transfer matrix, $\ket{i/f}$ are the initial/final states, $\ket{n}$ are the intermediate states. By the cutting equation \cite{Veltman}, the cut diagram corresponding to that in Fig.1 is shown in Fig.2 (Note that momentum flows from the shaded to the unshaded side of the cut), and this diagram must corresponds to terms on the R.H.S. of $(1)$ in order for unitarity to be satisfied. Since causality is imposed via the external wave packets, the cut diagram does not vanished. But because the neutrino is exchange through the t-channel, this diagram could not represent any of the terms on the R.H.S. of $(1)$, hence violating unitarity.  

This implies that field theoretical approaches to neutrino oscillation employing the exchange of virtual neutrinos is inherently inconsistent, and the probabilties thus derived may appear unitary only because they are normalized by hand. In general, the only meaningful temporal ordering in $S$-matices are between the $in/out$ states and not the interaction points, which exist in a cloud of space-time uncertainty.  

Since, as shown in \cite{Beuthe}, such a virtual neutrino model, with appropriately chosen external wave packets could reproduce the results of most neutrino oscillation models (even those that do not involve virtual neutrinos), it is therefore prudent to re-examine the neutrino oscillation probabilities, putting special emphasis on the issues of unitarity and causality. With these considerations in mind we will proceed to construct a field theoretical model of neutrino oscillation that will be free from the defects highlighted here.   

\begin{figure}[h]
\centering
\includegraphics[scale=0.6]{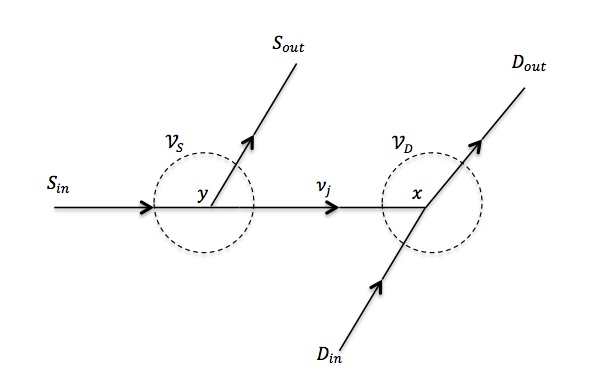}
\caption{\label{fig:fig1} Neutrino exchange amplitude. $\nu_{j}$ is a neutrino of mass $m_{j}$. $y/x$ are the space-time coordinates of the production/detection vertices, $\mathcal{V}_{S/D}$ are the corresponding space-time uncertainties}
\end{figure}

\begin{figure}[h]
\centering
\includegraphics[scale=0.6]{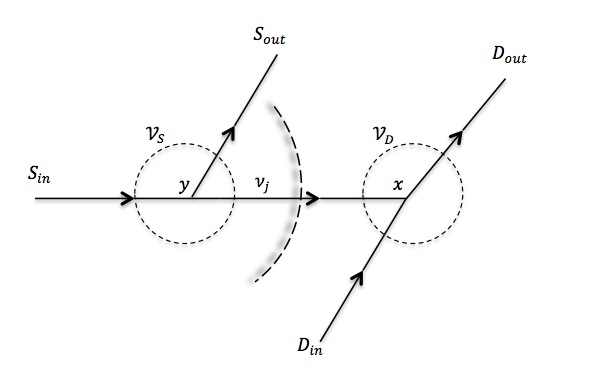}
\caption{\label{fig:fig2} Neutrino line is cut. Momentum flows from the shaded to unshaded side.}
\end{figure}

\section{An alternative model}

For the sake of generality, we assume the following neutrino producing decay
\begin{equation}
A \longrightarrow\nu_{\alpha} + B
\end{equation}
with $A$ as the unstable source particle, $\alpha$ as the initial flavor of the neutrino, $\nu_{\alpha}$, and $B$ is the decay products inclusive of the charged lepton, $l_{\alpha}$.

The neutrino detection process is given by
\begin{equation}
\nu_{\beta}+D \longrightarrow l_{\beta} + C
\end{equation}
where $D$ is the detector particle, $\beta$ is the flavor of the charge lepton produced from neutrino detection, and $C$ is the remaining products of this process. 

The interaction governing these 2 processes is taken to be local and can be describe by some Hamiltonian density, $\mathcal{H}^{(S/D)}_{int}(x)$ (which is assumed to be first order in the Fermi constant, $G_{F}$), where the superscript indicates source/detection, which may involve different interactions. Working in the Schrodinger picture, the time evolution operator describing neutrino production and detection is given by
\begin{equation}
U(t)=e^{-iH_{0}t}S(t)
\end{equation}
where $t$ is the interaction time of the processes, $H_{0}$ is the Hamiltionian describing the free propagation of particles in vacuum, and $S(t)$ is the scattering matrix
\begin{equation}
S(t)=Texp\Bigg(-i\int\limits_{0}^{t}dt'\int d^{3}x \mathcal{H}^{(S/D)}_{int}(\vec{x},t')\Bigg)
\end{equation} 
(the effects on neutrino oscillation due to the finite time of its production and detection is also given in \cite{Jun Wu}, using a planewave treatment. See also \cite{Grimus2000} for virtual neutrino oscillation with an unstable source).

    As previously stated, wave packets will be assigned only to the source particle and the detector particle (in-going states of interactions) and these are (We follow the convention of \cite{Akhmedov,Beuthe})
\begin{equation}
\ket{A}=\int [d\vec{p}_{A}]\Psi_{A}(\vec{p}_{A})\ket{\vec{p}_{A}}
\end{equation}
for the source, with $[d\vec{p}_{A}]\equiv d^{3}p_{A}(2\pi)^{-3}(2E_{A})^{-\frac{1}{2}}$, $E_{A}=\sqrt{\vec{p}_{A}^{2}+m^{2}_{A}}$. $\Psi_{A}(\vec{p}_{A})$ is the wave packet of $A$, localizing it within the source. The following set of normalizing conditions holds
\begin{equation}
\braket{\vec{p}_{A}|\vec{p}_{A} '}=(2\pi)^{3}2E_{A}\delta^{3}(\vec{p}_{A}-\vec{p}_{A} ')
\end{equation}
\begin{equation}
\int d^{3}p_{A}(2\pi)^{-3}\abs{\Psi_{A}}^{2}=1
\end{equation}
\begin{equation}
\braket{A|A}=1
\end{equation}

Similarly, for the detector particle, $D$
\begin{equation}
\ket{D}=\int [d\vec{p}_{D}]\Psi_{D}(\vec{p}_{D})e^{-i\vec{p}_{D}\cdot\vec{L}}\ket{\vec{p}_{D}}
\end{equation}
where $\Psi_{D}(\vec{p}_{D})$ localized $D$ to within the source, and the translation $e^{-i\vec{p}_{D}\cdot\vec{L}}$ shifts it across the macroscopic displacement $\vec{L}$, into the detector. Thus $\ket{D}$ is located at $\vec{L}$ from $\ket{A}$.

During the weak decay of $A$, its state evolves such that at $T_{1}$
\begin{equation}
\ket{A}\longrightarrow\Phi_{sur}(T_{1})\ket{A}+\Phi_{decay}(T_{1})\ket{\nu_{\alpha},l_{\alpha}}
\end{equation}
$\Phi_{sur}(T_{1})$ and $\Phi_{decay}(T_{1})$ are the survival and decay amplitude at $T_{1}$. $T_{1}$ is the time scale associated with the decoherence of state $(11)$ within the source, through interactions with the environment. This could occur via collisions between $A$ particles within the source or similar electromagnetic collisions, experienced by $l_{\alpha}$ ($\nu_{\alpha}$ is assumed to escape the source and only interacts with the detector). These predominantly electromagnetic collisional monitoring by the environment is assumed to occur much faster than the weak process of the decay; hence it makes sense to speak of the start and end of the interval $T_{1}$.  

The state $(11)$ collapsed either into the first term, at which the process is reset, and evolves until sufficient decoherence effects build up again at the end of the next interval $T_{1}$, or collapsed into the second term, whereby the decay occurs and the neutrino is emitted. $T_{1}$ can thus be understood as the time interval between incidents of environment monitoring beginning with an undecayed $A$ and ending with its decay. It is apparent that this second term is sufficient for one to construct the state of emitted neutrinos. The neutrinos are assumed to be external states and on mass-shell to avoid the problems discussed in Section II.

The transfer matrix element corresponding to (2) is given by
\begin{equation}
\bra{(\vec{p}_{k},s),\{\vec{p}_{B}\} }iT^{(S)}\ket{\vec{p}_{A}}\equiv U_{\alpha k}^{\ast}iT^{(S)}_{B}(\vec{p}_{k},s;\vec{p}_{A})(2\pi)^{3}\delta^{3}(\vec{p}_{A}-\vec{p}_{f})\int\limits^{T_{1}}_{0}dte^{-i(E_{A}-E_{f})t}
\end{equation}
where the convention, $S=1+iT$ is adopted for the definition of the transfer matrix. $(\vec{p}_{k},s)$ refers to the momentum of the neutrino of mass $m_{k}$, with helicity $s$, and $\{\vec{p}_{B}\}$ labels the momentum states of the set of particles $B$ (possible discrete quantum numbers are suppressed for simplicity). $\vec{p}_{B}$ is the total momentum of $B$, $\vec{p}_{f}=\vec{p}_{k}+\vec{p}_{B}$ and $E_{f}=E_{k}+E_{B}$ ($E_k=\sqrt{\vec{p}^{2}+m_{k}^{2}}$), $E_{B}$ is the total energy of $B$). The dependence on the unitary neutrino mixing matrix $U_{\alpha k}$, is factored out explicitly. And because we are working to the first order of $G_{F}$ (at the amplitude level), for a finite time evolution, the usual energy conserving delta function is replaced by a time integral.

$(12)$ can be rewritten as
\begin{eqnarray}
\bra{(\vec{p}_{k},s),\{\vec{p}_{B}\}}iT^{(S)}\ket{\vec{p}_{A}}\equiv \lefteqn{ U_{\alpha k}^{\ast} \int d\Delta_{1} \frac{e^{-i\Delta_{1}T_{1}/2}}{\pi \Delta_{1}}sin \Big(\frac{\Delta_{1} T_{1}}{2}\Big) } \nonumber \\
&& \cdot iT^{(S)}_{B}(\vec{p}_{k},s;\vec{p}_{A})(2\pi)^{4}\delta^{3}(\vec{p}_{A}-\vec{p}_{f})\delta(E_{A}-E_{f}-\Delta_{1})
\end{eqnarray}
where we have used
\begin{equation}
\int\limits^{T_{1}}_{0}dte^{-i(E_{A}-E_{f})t}=\int d\Delta_{1} \frac{e^{-i\Delta_{1}T_{1}/2}}{\pi \Delta_{1}}sin \Big(\frac{\Delta_{1} T_{1}}{2}\Big)2\pi \delta(E_{A}-E_{f}-\Delta_{1})
\end{equation}
Allowing us to reintroduced the usual energy conserving delta function, with the uncertainty controlled by the function $\Delta_{1}^{-1} sin(\frac{\Delta_{1} T_{1}}{2})$. 

Assuming that only neutrinos traveling along $\vec{L}$ ($\vec{L}=L\hat{x}$) are detected, the energy-momentum constraints imposed by the delta functions are 
\begin{equation}
\vec{p}_{A}=p_{k}\hat{x}+\vec{p}_{B}, E_{A}=E_{k}+E_{B}+\Delta_{1}
\end{equation}
$(15)$ could then be solved for $\vec{p}_{k}(\{\vec{p}_{B} \},\Delta_{1})$ and $\vec{p}_{A}(\{\vec{p}_{B} \},\Delta_{1})$. With $(4)$ and $(6)$, the joint $\nu-B$ state after decay is given by
\begin{equation}
\ket{\Psi_{\nu \otimes B}}\propto \int [d\vec{p}_{A}]\Psi_{A}(\vec{p}_{A})\int D_{\nu \otimes B}\ket{(\vec{p}_{k},s),\{\vec{p}_{B}\}}\bra{(\vec{p}_{k},s),\{\vec{p}_{B}\}}e^{-iH_{0}T_{1}} iT^{(S)}\ket{\vec{p}_{A}}
\end{equation}
$\int D_{\nu \otimes B}$ is the sum over the relevant final $\nu$-$B$ state (over $\hat{x}$ propagating neutrino). Performing the momentum integral $\int d^{3}p_{A}d^{3}p_{k}$, using $(13)$ and $(15)$, $(16)$ becomes
\begin{eqnarray}
\nonumber \ket{\Psi_{\nu \otimes B}}\propto && \int D_{B}e^{-iE_{B}T_{1}}\sum_{k,s} U^{\ast}_{\alpha k} \int d\Delta_{1}e^{-i(\Delta_{1}T_{1}/2+E_{k}(\vec{p}_{k}(\{\vec{p}_{B}\},\Delta_{1}))T_{1})}\frac{1}{\Delta_{1}}sin\Big(\frac{\Delta_{1} T_{1}}{2}\Big)  \\ \nonumber
 && \cdot \frac{1}{(2\pi)^{3}\sqrt{2E_{A}(\vec{p}_{A}(\{\vec{p}_{B}\},\Delta_{1}))}} \cdot \frac{1}{(2\pi)^{3}2E_{k}(\vec{p}_{k}(\{\vec{p}_{B}\},\Delta_{1}))}  \\ 
&& \cdot iT^{(S)}_{B}(\vec{p}_{k}(\{\vec{p}_{B}\},\Delta_{1}),s;\vec{p}_{A}(\{\vec{p}_{B}\},\Delta_{1})) \cdot \Psi (\vec{p}_{A}(\{\vec{p}_{B}\},\Delta_{1}))\ket{(\vec{p}_{k}(\{\vec{p}_{B}\},\Delta_{1}),s),\{\vec{p}_{B}\}}
\end{eqnarray}

The factor $(2\pi)^{-3}(2E_{k}(\vec{p}_{k}(\{\vec{p}_{B}\},\Delta_{1})))^{-1}$ come from the neutrino phase space factor of the integral, $\int D_{\nu}$. In the next section, we shall derive the neutrino state from $\ket{\Psi_{\nu \otimes B}}$.

\subsection{The neutrino state}

To obtain the neutrino state right after emission, we need to trace out the $B-state$ from $\ket{\Psi_{\nu \otimes B}}$. This is easily done using $(17)$
\begin{equation}
\rho_{\nu}=tr\{\ket{\Psi_{\nu \otimes B}}\bra{\Psi_{\nu \otimes B}}\}=\int D_{B} g(\{\vec{p}_{B}\})\ket{\nu_{\alpha}(T_{1},\{\vec{p}_{B}\})}\bra{\nu_{\alpha}(T_{1},\{\vec{p}_{B}\})}
\end{equation}
where $g(\{\vec{p}_{B}\})$ is a probability distribution of the classical sum of the neutrino state $\ket{\nu_{\alpha}(T_{1},\{\vec{p}_{B}\})}$, which is the state that occurs in a product with the momentum $B-state$ $\ket{\{\vec{p}_{B}\}}$ in $(17)$ that satisfies the orthonormality condition
\begin{equation}
\braket{\{\vec{p}_{B}\}|\{\vec{p}_{B}'\}} \propto \delta^{3}(\vec{p}_{B}-\vec{p}_{B}')
\end{equation}
where the proportionality sign indicates the we are comparing entire configurations of $B-states$ (not just their total momentum), and that a host of other quantum numbers associated with $B$ are not expressed explicitly. From $(17)$ this neutrino state is
\begin{equation}
\ket{\nu_{\alpha}(T_{1},\{\vec{p}_{B}\})}=\sum_{k} U^{\ast}_{\alpha k} \ket{\nu_{k}(T_{1},\{\vec{p}_{B}\})}
\end{equation}
with the neutrino mass eigenstate
\begin{eqnarray}
\nonumber \ket{\nu_{k}(T_{1},\{\vec{p}_{B}\})} \propto && \sum_{s} \int d\Delta_{1}e^{-i(\Delta_{1}T_{1}/2+E_{k}(\vec{p}_{k}(\{\vec{p}_{B}\},\Delta_{1}))T_{1})}\frac{1}{\Delta_{1}}sin\Big(\frac{\Delta_{1} T_{1}}{2}\Big)  \\ \nonumber
 && \cdot \frac{1}{(2\pi)^{3}\sqrt{2E_{A}(\vec{p}_{A}(\{\vec{p}_{B}\},\Delta_{1}))}} \cdot \frac{1}{(2\pi)^{3}2E_{k}(\vec{p}_{k}(\{\vec{p}_{B}\},\Delta_{1}))}  \\ \nonumber
&& \cdot iT^{(S)}_{B}(\vec{p}_{k}(\{\vec{p}_{B}\},\Delta_{1}),s;\vec{p}_{A}(\{\vec{p}_{B}\},\Delta_{1})) \\
&& \cdot \Psi (\vec{p}_{A}(\{\vec{p}_{B}\},\Delta_{1}))\ket{\vec{p}_{k}(\{\vec{p}_{B}\},\Delta_{1}),s}
\end{eqnarray} 
Note that this is a neutrino wave packet formed only because of the $\Delta_{1}$-dependence of the neutrino momentum $\vec{p}_{k}(\{\vec{p}_{B}\},\Delta_{1})$, which is the result of energy uncertainty due to the finite time of decay of the parent, $A$.  
No requirement is made on how the $B-state$s are measured or if they were measured. The basis, $\ket{ \{ \vec{p}_{B} \} }$, in which $B$ is traced out of the decayed state, $\ket{\Psi_{\nu \otimes B}}$, is chosen purely out of convenience.

After the neutrino is emitted, it propagates freely for the time, $T_{prop}$
\begin{equation}
\ket{\nu_{k}(T_{1}+T_{prop},\{\vec{p}_{B}\})}=e^{-iH_{0}T_{1}} \ket{\nu_{k}(T_{1},\{\vec{p}_{B}\})}
\end{equation}
before the starting to interact with the detector particle, $D$. From $(21)$ and $(22)$
\begin{eqnarray}
\nonumber \ket{\nu_{k}(T_{1}+T_{prop},\{\vec{p}_{B}\})} \propto && \sum_{s} \int d\Delta_{1}e^{-i(\Delta_{1}T_{1}/2+E_{k}(\vec{p}_{k}(\{\vec{p}_{B}\},\Delta_{1}))(T_{1}+T_{prop}))}\frac{1}{\Delta_{1}}sin\Big(\frac{\Delta_{1} T_{1}}{2}\Big)  \\ \nonumber
 && \cdot \frac{1}{(2\pi)^{3}\sqrt{2E_{A}(\vec{p}_{A}(\{\vec{p}_{B}\},\Delta_{1}))}} \cdot \frac{1}{(2\pi)^{3}2E_{k}(\vec{p}_{k}(\{\vec{p}_{B}\},\Delta_{1}))}  \\ \nonumber
&& \cdot iT^{(S)}_{B}(\vec{p}_{k}(\{\vec{p}_{B}\},\Delta_{1}),s;\vec{p}_{A}(\{\vec{p}_{B}\},\Delta_{1})) \\
&& \cdot \Psi (\vec{p}_{A}(\{\vec{p}_{B}\},\Delta_{1}))\ket{\vec{p}_{k}(\{\vec{p}_{B}\},\Delta_{1}),s}
\end{eqnarray} 
Writing this in a normalized form
\begin{eqnarray}
\nonumber \lefteqn{ \ket{\nu_{k}(T_{1}+T_{prop},\{\vec{p}_{B}\})} } \\
&& = \sum_{s} \int d\Delta_{1} \tilde{\Psi}_{k,B}(\Delta_{1}) e^{-i(\Delta_{1}T_{1}/2+E_{k}(\vec{p}_{k}(\{\vec{p}_{B}\},\Delta_{1}))(T_{1}+T_{prop}))} \nonumber \\
&& \cdot sinc\Big(\frac{\Delta_{1} T_{1}}{2}\Big) \ket{(\vec{p}_{k}(\{\vec{p}_{B}\},\Delta_{1}),s}
\end{eqnarray} 
where we have introduced the function $sinc(x)\equiv sin(x) \slash x$,
\begin{eqnarray}
\nonumber \lefteqn{ \tilde{\Psi}_{k,B}(\Delta_{1}) } \\
&& =\frac{1}{\mathcal{N}_{k}(\{\vec{p}_{B}\})} \cdot \frac{\sqrt{E_{A}(\vec{p}_{A}(\{\vec{p}_{B}\},\Delta_{1}))}}{\abs{(p_{B,x}+p_{k}(\{\vec{p}_{B}\},\Delta_{1}))E_{k}(\vec{p}_{k}(\{\vec{p}_{B}\},\Delta_{1}))-p_{k}(\{\vec{p}_{B}\},\Delta_{1})E_{A}(\vec{p}_{A}(\{\vec{p}_{B}\},\Delta_{1}))}} \nonumber \\
&& \cdot \Psi_{A} (\vec{p}_{A}(\{\vec{p}_{B}\},\Delta_{1})) \cdot  iT^{(S)}_{B}(\vec{p}_{k}(\{\vec{p}_{B}\},\Delta_{1}),s;\vec{p}_{A}(\{\vec{p}_{B}\},\Delta_{1}))
\end{eqnarray} 
and the normalizing factor is,
\begin{eqnarray}
\nonumber \lefteqn{ \mathcal{N}_{k}(\{\vec{p}_{B}\}) } \\
&& = \Bigg[ \sum_{s} \int d\Delta_{1}  \frac{16\pi^{3}}{\abs{(p_{B,x}+p_{k}(\{\vec{p}_{B}\},\Delta_{1}))E_{k}(\vec{p}_{k}(\{\vec{p}_{B}\},\Delta_{1}))-p_{k}(\{\vec{p}_{B}\},\Delta_{1})E_{A}(\vec{p}_{A}(\{\vec{p}_{B}\},\Delta_{1}))}} \nonumber \\
&& \cdot \abs{\Psi_{A} (\vec{p}_{A}(\{\vec{p}_{B}\},\Delta_{1}))}^{2} \cdot \abs{ iT^{(S)}_{B}(\vec{p}_{k}(\{\vec{p}_{B}\},\Delta_{1}),s;\vec{p}_{A}(\{\vec{p}_{B}\},\Delta_{1}))}^{2} sinc^{2}\Big(\frac{\Delta_{1} T_{1}}{2}\Big) \Bigg]^{1/2}
\end{eqnarray} 
such that,
\begin{equation}
\braket{\nu_{k}(T_{1}+T_{prop},\{\vec{p}_{B}\})|\nu_{k'}(T_{1}+T_{prop},\{\vec{p}_{B}\})}=\delta_{kk'}
\end{equation}
so that via $(20)$
\begin{equation}
\braket{\nu_{\alpha}(T_{1}+T_{prop},\{\vec{p}_{B}\})|\nu_{\alpha}(T_{1}+T_{prop},\{\vec{p}_{B}\})}=1
\end{equation}
$\ket{\nu_{\alpha}(T_{1}+T_{prop},\{\vec{p}_{B}\})}$ will be the in-going neutrino state for the detection amplitude.

\subsection{Neutrino detection amplitude}

In this section, we will derive the formal expression for the neutrino detection amplitude ($\mathcal{A}(\beta|\alpha,B;C,\vec{L},T_{prop})$) after the incident neutrino has been interacting with the detector particle, $D$, for time $T_2$. The amplitude for the process $(3)$ is given by
\begin{eqnarray}
\nonumber \lefteqn{ \mathcal{A}(\beta|\alpha,B;C,\vec{L},T_{prop}) } \\
&=& \bra{C\otimes l_{\beta}} e^{-iH_{0}T_{2}} \cdot iT^{(S)} \ket{\nu_{\alpha}(T_{1}+T_{prop},\{\vec{p}_{B}\}) \otimes D}  \nonumber \\
&=& \sum_{k,s} U^{\ast}_{\alpha k} \int d\Delta_{1} \tilde{\Psi}_{k,B}(\Delta_{1}) e^{-i(\Delta_{1}T_{1}/2+E_{k}(\vec{p}_{k}(\{\vec{p}_{B}\},\Delta_{1}))(T_{1}+T_{prop}))} \nonumber \\
&& \cdot sinc\Big(\frac{\Delta_{1} T_{1}}{2}\Big) \int [d \vec{p}_{D}] \Psi_{D}(\vec{p}_{D}) e^{-i\vec{p}_{D} \cdot \vec{L}} \bra{C\otimes l_{\beta}} e^{-iH_{0}T_{2}} \cdot iT^{(D)} \ket{(\vec{p}_{k}(\{\vec{p}_{B}\},\Delta_{1}),s) \otimes \vec{p}_{D}}
\end{eqnarray}
with the momentum space detection transfer matrix element
\begin{eqnarray}
\nonumber \lefteqn{ \bra{C\otimes l_{\beta}} e^{-iH_{0}T_{2}} \cdot iT^{(D)} \ket{(\vec{p}_{k}(\{\vec{p}_{B}\},\Delta_{1}),s) \otimes \vec{p}_{D}} } \\
&=& e^{i(E_{C}+E_{l_{\beta}})T_{2}} \cdot U_{\beta k} \cdot iT^{(D)}_{C}\big((\vec{p}_{l_{\beta}},s_{l_{\beta}});(\vec{p}_{k}(\{\vec{p}_{B}\},\Delta_{1}),s), \vec{p}_{D} \big) \nonumber \\
&& \cdot (2\pi)^{3} \delta^{3}(\vec{p}_{f'}-\vec{p}_{k}(\{\vec{p}_{B}\},\Delta_{1})-\vec{p}_{D}) \nonumber \\
&& \cdot T_{2} e^{-i(E_{k}(\vec{p}_{k}(\{\vec{p}_{B}\},\Delta_{1}))+E_{D}-E_{f'})T_{2}/2} sinc\big( ( E_{k}(\vec{p}_{k}(\{\vec{p}_{B}\},\Delta_{1}) )+E_{D}-E_{f'})T_{2}/2 \big) 
\end{eqnarray}
and we have used $(10)$ and $(24)$. $\vec{p}_{f'}=\vec{p}_{C}+\vec{p}_{l_{\beta} }$, $E_{f'}=E_{C}+E_{l_{\beta}}$, and $l_{\beta}$ labels quantities associated with the charge lepton produced during detection. $E_{C}$ is the total energy of $C$. 

The phase, $e^{i(E_{C}+E_{l_{\beta} } ) T_{2} }$, in $(30)$ can be dropped as it corresponds to the final state of the detection process and is only an overall phase to the amplitude.

   Substituting $(30)$ into $(29)$ and performing $\int[d\vec{p}_{D} ]$, we have
\begin{eqnarray}
\nonumber \lefteqn{ \mathcal{A}(\beta|\alpha,B;C,\vec{L},T_{prop}) } \\
&=&T_{2} \sum_{k} U^{\ast}_{\alpha k} U_{\beta k} \sum_{s} \int d\Delta_{1} \tilde{\Psi}_{k,B}(\Delta_{1}) e^{-i(\Delta_{1}T_{1}/2+E_{k}(\vec{p}_{k}(\{\vec{p}_{B}\},\Delta_{1}))(T_{1}+T_{prop}))} \nonumber \\
&& \cdot sinc\Big(\frac{\Delta_{1} T_{1}}{2}\Big) \nonumber \\
&& \cdot \Bigg[ \frac{1}{\sqrt{2E_{D}}} \Psi_{D}(\vec{p}_{D}) e^{-i\vec{p}_{D} \cdot \vec{L}} \cdot iT^{(D)}_{C}\big((\vec{p}_{l_{\beta}},s_{l_{\beta}});(\vec{p}_{k}(\{\vec{p}_{B}\},\Delta_{1}),s), \vec{p}_{D} \big) \nonumber \\
&& \cdot e^{-i(E_{k}(\vec{p}_{k}(\{\vec{p}_{B}\},\Delta_{1}))+E_{D}-E_{f'})T_{2}/2} \nonumber \\
&& \cdot sinc\big( (E_{k}(\vec{p}_{k}(\{\vec{p}_{B}\},\Delta_{1}))+E_{D}-E_{f'})T_{2}/2 \big) \Bigg]_{\vec{p}_{D}=\vec{p}_{f'}-\vec{p}_{k}(\{\vec{p}_{B}\},\Delta_{1})}
\end{eqnarray}
The replacement, $\vec{p}_{D}=\vec{p}_{f'}-\vec{p}_{k}(\{\vec{p}_{B}\},\Delta_{1})$, is due to the momentum delta function in $(30)$. To evaluate $(31)$ and subsequently the detection probability, we have to make some approximations. This is the subject of the next section.

\subsection{Neutrino spatial wave packet}

In this section, we shall solve for the spatial wave packet corresponding the neutrino state given in $(24)$. 

   The neutrino spatial wave function (for mode $k$) is defined to be (helicity is neglected)  
\begin{equation}
\psi_{k,s}(x,T_{1}+T_{prop}) \equiv \braket{x|\nu_{k}(T_{1}+T_{prop})},
\end{equation}
where we have used,
\begin{eqnarray}
\lefteqn{\psi_{k,s}(x,T_{1}+T_{prop})} \nonumber \\
&&= \int d\Delta_{1} (2\pi)^{3/2} \sqrt{2E_{k}(\vec{p}_{k}(\Delta_{1}))} \tilde{\Psi}_{k,B}(\Delta_{1}) e^{-i\Delta_{1}T_{1}/2} e^{i \big( p_{k}(\Delta_{1})x-E_{k}(\vec{p}_{k}(\Delta_{1}))(T_{1}+T_{prop}) \big) } \nonumber \\
&& \cdot sinc\Big( \frac{\Delta_{1}T_{1}}{2} \Big)
\end{eqnarray}
where we have used,
\begin{equation}
\braket{x|\vec{p}_{k}(\Delta_{1})}=(2\pi)^{3/2} \sqrt{2E_{k}(\vec{p}_{k}(\Delta_{1}))} e^{ip_{k}(\Delta_{1})x}
\end{equation}
which is consistent with the normalization in $(7)$. Applying the approximations in the appendix, and using the identities (A.12) and (A.13), we have
\begin{equation}
\psi_{k,s}(x,T_{1}+T_{prop}) \propto rect \Big[ \frac{2}{T_{1} \abs{v_{A,x}-v_{k}}} \Big( x-\big( v_{k}(T_{1}+T_{prop}) +\frac{T_{1}}{2} (v_{A,x}-v_{k}) \big) \Big) \Big]
\end{equation}
assuming that (which will be adopted hereafter),
\begin{equation}
v_{k} > \abs{v_{A,x}}, \abs{v_{D,x}}
\end{equation}
the wave packet can be expressed as
\begin{equation}
\psi_{k,s}(x,T_{1}+T_{prop}) \propto rect \Big[ \frac{2}{ \delta x_{\nu}} \Big( x-\big( v_{k}(T_{1}+T_{prop}) +\frac{\delta x_{\nu}}{2} \big) \Big) \Big]
\end{equation}
with its width given by,
\begin{equation}
\delta x_{\nu}=T_{1} \abs{v_{A,x}-v_{k}}
\end{equation}

We can see from $(37)$ that $v_{k} (T_{1}+T_{prop} )$ is the position of the leading edge (wave front) of the neutrino wave packet at time $t=T_{1}+T_{prop}$, from a spatial origin defined by the position of the source particle $A$ at $t=0$. It is also apparent from $(38)$ that the wave packet does not spread with $T_{prop}$, and only develops with $T_{1}$, before decoherence occurs between the decayed and undecayed $A$ state. The neutrino wave front propagates at the velocity $v_{k}$ both during the development of the wave packet and during propagation. 

   One could check the condition of zero neutrino detection amplitude $(A.14)$ using this spatial wave packet picture, by assuming that weak interaction between particles only occurs when their wave functions overlap, and that the detection amplitude will be zero when after $t=T_{1}+T_{prop}+T_2$ (recall that $T_{2}$ is the interaction time between the detector particle $D$ and the neutrino) if no instances of wave function overlap have occurred during the time interval $T_{2}$ (which begins right after $t=T_{1}+T_{prop}$). This is illustrated in Figs. (3 and 4).   

   By the assumptions in Section II.C, the spatial spread of $D$ is negligible compared to the neutrino ($\delta x_{\nu} \gg \delta x_{D}$, as $ \delta x_{D} \ll T_{1} $ and $v_{k} \sim 1$), as presented in both figures.

    Fig. 3 shows the lower limit of $v_{k} (T_{1}+T_{prop} )$ at $t=T_{1}+T_{prop}$. One could see that at $t=T_{1}+T_{prop}+T_{2}$, the gap between the neutrino and detector particle wave packets ($\psi_{k,s}$ and $\psi_{D}$), $T_{2} \abs{v_{k}-v_{D,x}}$, will be just closed if $D$ is at position $L$ at $t=T_{1}+T_{prop}$. Hence below this limit no wave functions overlap and therefore, no absorption (contact interaction assumption) is possible by $t=T_{1}+T_{prop}+T_{2}$. This corresponds to the first condition for $\mathcal{A}(\beta|\alpha,B;C,\vec{L},T_{prop})=0$, in (A.14).

    Fig. 4 shows the upper bound of $v_{k} (T_{1}+T_{prop} )$ at time, $t=T_{1}+T_{prop}$. As can be seen from the figure, since $v_{k} >\abs{v_{D,x}}$, beginning with an initial non-overlapping borderline case, with the trailing edge of the wave function of the neutrino just ahead of $D$, the latter can never catch up with the former in the time interval $T_{2}$. This corresponds to the second condition for $\mathcal{A}(\beta|\alpha,B;C,\vec{L},T_{prop})=0$, in (A.14).    

   These considerations lead exactly to the conditions of non-detection of the neutrino in (A.14), hence providing a consistent physical picture to the amplitude, $\mathcal{A}(\beta|\alpha,B;C,\vec{L},T_{prop})$. In Section II.D, we shall proceed with the evaluation of the non-zero neutrino detection amplitude and the corresponding probability, employing the ideas developed in this section (local weak interaction of wave packets).   

\begin{figure}[h]
\centering
\includegraphics[scale=0.6]{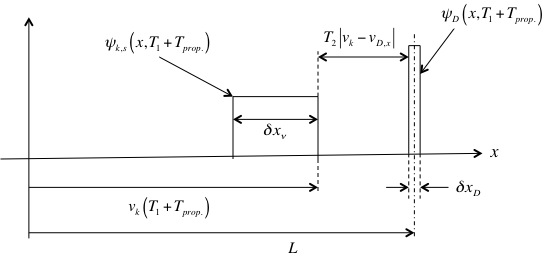}
\caption{\label{fig:fig3} Shown is the lower limit of $v_{k} (T_{1}+T_{prop})$ at $t=T_{1}+T_{prop}$, below which no overlap (hence no absorption) between $\psi_{k,s}$  and $\psi_{D}$ is possible by the time $t=T_{1}+T_{prop}+T_{2}$.}
\end{figure}

\begin{figure}[h]
\centering
\includegraphics[scale=0.6]{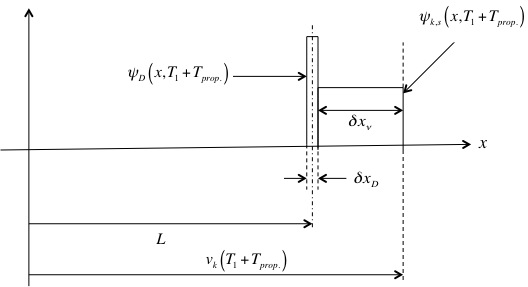}
\caption{\label{fig:fig4} Shown is the upper limit of $v_{k} (T_{1}+T_{prop})$ at $t=T_{1}+T_{prop}$, above which no wave functions overlap is possible within the time interval $T_{2}$}
\end{figure}

\subsection{Non-zero neutrino detection amplitude and probability}
From (A.4) using (A.5), (A.8) and (A.10), the non-zero neutrino detection amplitude can be expressed as 
\begin{eqnarray}
\nonumber \lefteqn{ \mathcal{A}(\beta|\alpha,B;C,\vec{L},T_{prop}) } \\
&\simeq&T_{2} \sum_{k} U^{\ast}_{\alpha k} U_{\beta k} \sum_{s}  \frac{1}{\sqrt{2E_{D}}} \tilde{\Psi}_{k,B}(\Delta_{1}=0) \Psi_{D}(\vec{p}_{D}) e^{-i\vec{p}_{f'} \cdot \vec{L}} e^{iE_{f'}T_{2}/2}  \nonumber \\ 
&& \cdot iT^{(D)}_{C}\big((\vec{p}_{l_{\beta}},s_{l_{\beta}});(\vec{p}_{k},s), \vec{p}_{D} \big) \Big( -\frac{2\pi}{T_{1}} \Big) e^{i\phi_{k}} \frac{sin(\Theta_{k})}{(E_{k}+E_{D}-E_{f'})T_{2}/2}; \nonumber \\
&& \Theta_{k}=\frac{E_{k}+E_{D}-E_{f'}}{2(v_{k}-v_{D,x})} \chi_{k}, \chi_{k}=(v_{A,x}-v_{k})(k_{2}-k_{1}), \nonumber \\
&& \phi_{k}=\Big[  -E_{k}(T_{1}+T_{prop})+p_{k}L-(E_{k}+E_{D}) \frac{T_{2}}{2} \Big] +\frac{E_{k}+E_{D}-E_{f'}}{2(v_{k}-v_{D,x})} \cdot \omega_{k}
\end{eqnarray}
where we have applied (A.5) and (A.10) to $\phi_{k} \equiv G-\frac{cF}{b}+\frac{c}{2b} (k_{1}+k_{2} )$. The expressions for $\omega_{k}$ and $\chi_{k}$ are collected in Table (I and II), classified according to the six cases in (A.9) (see also Fig. (5 to 10)).

\begin{table}[h]
\centering
\begin{tabular}{| r | c | c | c |}
\hline
Case & $\omega_{k}$ & Domain & Condition \\
\hline\hline
a.) & $v_{k}(T_{1}+T_{prop})-L$                                             & $0<v_{k}(T_{1}+T_{prop})-L<\abs{v_{A,x}-v_{k}} T_{1}$  &             \\  
     & $+\abs{v_{k}-v_{D,x}} T_{2}-\abs{v_{A,x}-v_{k}} T_{1}$ &                                                                                           &             \\  
\cline{1-3}
b.) & $2(v_{k}(T_{1}+T_{prop})-L)$                                         & $\abs{v_{A,x}-v_{k}} T_{1} -\abs{v_{k}-v_{D,x}} T_{2} <$ &  $\abs{v_{k}-v_{D,x}} T_{2}$\\  
     & $+\abs{v_{k}-v_{D,x}} T_{2}-\abs{v_{A,x}-v_{k}} T_{1}$  & $v_{k}(T_{1}+T_{prop})-L< 0$                                          & $>\abs{v_{A,x}-v_{k}} T_{1}$ \\  
\cline{1-3}
c.) & $v_{k}(T_{1}+T_{prop})-L$                                              & $ -\abs{v_{k}-v_{D,x}} T_{2} <v_{k}(T_{1}+T_{prop})-L<$ &              \\  
     &                                                                                         & $ \abs{v_{A,x}-v_{k}} T_{1}-\abs{v_{k}-v_{D,x}} T_{2}$     &               \\  
\hline
d.) & $v_{k}(T_{1}+T_{prop})-L$                                             & $\abs{v_{A,x}-v_{k}} T_{1} -\abs{v_{k}-v_{D,x}} T_{2}< $      &             \\  
     & $+\abs{v_{k}-v_{D,x}} T_{2}-\abs{v_{A,x}-v_{k}} T_{1}$ &  $v_{k}(T_{1}+T_{prop})-L<\abs{v_{A,x}-v_{k}} T_{1}$        &             \\  
\cline{1-3}
e.) & 0                                                                                      & $0<v_{k}(T_{1}+T_{prop})-L< $                                         &  $\abs{v_{k}-v_{D,x}} T_{2}$\\  
     &                                                                                          & $\abs{v_{A,x}-v_{k}} T_{1}$                                               & $<\abs{v_{A,x}-v_{k}} T_{1}$ \\  
\cline{1-3}
f.) & $v_{k}(T_{1}+T_{prop})-L$                                              & $ -\abs{v_{k}-v_{D,x}} T_{2} <v_{k}(T_{1}+T_{prop})-L<0$ &              \\  
     &                                                                                         &                                                                                              &               \\  
\hline
\end{tabular}
\caption{Values of $\omega_{k}$ corresponding to the domains of $v_{k} (T_{1}+T_{prop} )-L$.}
\end{table}

\begin{table}[h]
\centering
\begin{tabular}{| r | c | c | c |}
\hline
Case & $\chi_{k}$ & Domain & Condition \\
\hline\hline
a.) & $v_{k}(T_{1}+T_{prop})-L$                                             & $0<v_{k}(T_{1}+T_{prop})-L<\abs{v_{A,x}-v_{k}} T_{1}$  &             \\  
     & $+\abs{v_{A,x}-v_{k}} T_{1}$                                          &                                                                                           &             \\  
\cline{1-3}
b.) & $(v_{A,x}-v_{k}) T_{1}$                                                   & $\abs{v_{A,x}-v_{k}} T_{1} -\abs{v_{k}-v_{D,x}} T_{2} <$ &  $\abs{v_{k}-v_{D,x}} T_{2}$\\  
     &                                                                                         & $v_{k}(T_{1}+T_{prop})-L< 0$                                          & $>\abs{v_{A,x}-v_{k}} T_{1}$ \\  
\cline{1-3}
c.) & $-(v_{k}(T_{1}+T_{prop})-L$                                           & $ -\abs{v_{k}-v_{D,x}} T_{2} <v_{k}(T_{1}+T_{prop})-L<$ &              \\  
     & $+\abs{v_{k}-v_{D,x}} T_{2})$                                          & $ \abs{v_{A,x}-v_{k}} T_{1}-\abs{v_{k}-v_{D,x}} T_{2}$     &               \\  
\hline
d.) & $v_{k}(T_{1}+T_{prop})-L$                                             & $\abs{v_{A,x}-v_{k}} T_{1} -\abs{v_{k}-v_{D,x}} T_{2}< $      &             \\  
     & $-\abs{v_{A,x}-v_{k}} T_{1}$                                           &  $v_{k}(T_{1}+T_{prop})-L<\abs{v_{A,x}-v_{k}} T_{1}$        &             \\  
\cline{1-3}
e.) &$-\abs{v_{k}-v_{D,x}} T_{2} $                                           & $0<v_{k}(T_{1}+T_{prop})-L< $                                         &  $\abs{v_{k}-v_{D,x}} T_{2}$\\  
     &                                                                                          & $\abs{v_{A,x}-v_{k}} T_{1}$                                               & $<\abs{v_{A,x}-v_{k}} T_{1}$ \\  
\cline{1-3}
f.) & $-(v_{k}(T_{1}+T_{prop})-L$                                              & $ -\abs{v_{k}-v_{D,x}} T_{2} <v_{k}(T_{1}+T_{prop})-L<0$ &              \\  
    & $+\abs{v_{k}-v_{D,x}} T_{2})$                                          &                                                                                              &               \\  
\hline
\end{tabular}
\caption{Values of $\chi_{k}$ corresponding to the domains of $v_{k} (T_{1}+T_{prop} )-L$.}
\end{table}

Note that due to the assumption $(36)$, $sgn(b)=-1$ (see also (A.9) and (A.10)), and that momentum conservation ($\vec{p}_{D} = \vec{p}_{f'} -\vec{p}_{k}$) is implicit in $(39)$.

   The probability associated with the amplitude $(39)$, defined in (A.1), now takes the form
\begin{eqnarray}
\nonumber \lefteqn{ P(\beta|\alpha,B;\vec{L},T_{prop}) } \\
&\simeq& \Big( \frac{2\pi T_{2}}{T_{1}} \Big)^{2} \sum_{k,k'} U^{\ast}_{\alpha k} U_{\beta k} U_{\alpha k'} U^{\ast}_{\beta k'} \sum_{s,s'} \int D_{l_{\beta} \otimes C} \Big[ \frac{1}{\sqrt{2E_{D}}} \tilde{\Psi}_{k,B}(\Delta_{1}=0) \Psi_{D}(\vec{p}_{D})   \nonumber \\ 
&& \cdot iT^{(D)}_{C}\big((\vec{p}_{l_{\beta}},s_{l_{\beta}});(\vec{p}_{k},s), \vec{p}_{D} \big) \Big]  \nonumber \\
&& \cdot \Big[ k \rightarrow k', s \rightarrow s' \Big]^{\ast}  e^{i(\phi_{k}-\phi_{k'})} \frac{sin(\Theta_{k})}{(E_{k}+E_{D}^{(k)}-E_{f'})T_{2}/2} \cdot \frac{sin(\Theta_{k'})}{(E_{k'}+E_{D}^{(k')}-E_{f'})T_{2}/2}
\end{eqnarray}

where $[k \rightarrow k', s \rightarrow s']^{\ast}$ is the complex conjugate of the preceding bracket with the indicated substitutions, and the superscript, $(k)$ in $E_{D}^{(k)}$ indicates its dependence on neutrino masses through momentum conservation. 

   At this point, in order to facilitate comparisons with standard approaches to neutrino oscillation, we make the usual assumptions of the ultra-relativistic neutrino, which as consequences, lead to the amplitude associated with the left (right) helicity neutrino (anti-neutrino) being suppressed (hence allowing us to drop the summation over $s$ in $(40)$), and that any expansion of energies and momenta be sufficient to the first order in $m_{k}^{2}$.

   Furthermore, to be consistent with the earlier assumption of local weak interaction of the neutrino with other particles, detection interaction of the neutrino is set to begin when the wave front of the leading neutrino wave packet, which corresponds to the fastest (hence lightest) mass mode, starts to overlap with $D$, and ends when the trailing end of the slowest (hence heaviest) mass mode passes $D$. These give rise to the following conditions,
\begin{equation}
T_{1}+T_{prop}=L/v_{L}
\end{equation}
\begin{equation}
v_{H}(T_{1}+T_{prop}+T^{max}_{2})-\abs{v_{A,x}-v_{H}} T_{1} = L+v_{D,x} T^{max}_{2}
\end{equation}
where $v_{L \slash H}$ the velocity of the lightest/heaviest neutrino, $T_{2}^{max}$ is the upper limit of $T_{2}$. The L.H.S. of $(42)$ is the expression for the position of the trailing edge of the heaviest neutrino.

   $(42)$ can be rewritten as 
\begin{equation}
\abs{v_{H}-v_{D,x}}T^{max}_{2} =\abs{v_{A,x}-v_{H}} T_{1} + (v_{L}-v_{H})(T_{1}+T_{prop})
\end{equation}
which results in the inequality 
\begin{equation}
\abs{v_{H}-v_{D,x}}T_{2}<  \abs{v_{A,x}-v_{H}} T_{1} + \frac{\Delta m^{2}_{HL}}{2E^{2}}(T_{1}+T_{prop})
\end{equation}
where we have used
\begin{equation}
\abs{v_{k}-v_{k'}} \simeq \abs{m^{2}_{kk'}\frac{\partial v_{k}}{m^{2}_{k}}} = \abs{\frac{\Delta m^{2}_{kk'}}{2E^{2}}\Big( \frac{1-v_{k}v_{A,x}}{v_{A,x}-v_{k}} \Big)} \simeq \abs{\frac{\Delta m^{2}_{kk'}}{2E_{k}^{2}}}
\end{equation}
and $E=(E_{k} )_{m_{k}^{2}=0}$ (velocity identities for the derivatives, $(\partial \slash \partial m_{k}^{2})_{m_{k}^{2}=0}$, of kinematic variables, similar to (A.12) and (A.13), can be derived from the energy-momentum conservation relations through a procedure analogous to (A.11)).
 
   From $(44)$, adding and subtracting the term $v_{k} T_{2}$ on the L.H.S., and $v_{k} T_{1}$ on the R.H.S. gives the condition
\begin{equation}
\abs{v_{k}-v_{D,x}}T_{2}<  \abs{v_{A,x}-v_{k}} T_{1} + \frac{\Delta m^{2}_{kH}}{2E^{2}}(T_{1}-T_{2})+\frac{\Delta m^{2}_{HL}}{2E^{2}}(T_{1}+T_{prop})
\end{equation}

If it is further assumed (which is consistent with ultra-relativistic neutrinos) that 
\begin{equation}
\frac{\Delta m^{2}_{HL}}{2E^{2}}(T_{1}+T_{prop}) \ll \abs{v_{A,x}-v_{k}} T_{1}, \abs{v_{k}-v_{D,x}}T_{2}
\end{equation}
$(46)$ reduces to
\begin{equation}
 \abs{v_{A,x}-v_{k}} T_{1} > \abs{v_{k}-v_{D,x}}T_{2}
\end{equation}
Thus we need only to consider cases d), e) and f) in Table (1 and 2). With approximations $(45)$ and $(47)$, $(41)$ can be taken to hold for any neutrino mass mode,
\begin{equation}
T_{1}+T_{prop}=L/v_{k}
\end{equation}
Applying $(47)$ and $(49)$ to $\Theta_{k}$ (for the cases d), e) and f)) we have
\begin{equation}
\Theta_{k} \simeq -\frac{(E_{k}+E_{D}^{(k)}-E_{f'})T_{2}}{2}
\end{equation}
This means that in $(40)$
\begin{equation}
\frac{sin(\Theta_{k})}{(E_{k}+E_{D}^{(k)}-E_{f'})T_{2}/2} \rightarrow \frac{sin((E_{k}+E_{D}^{(k)}-E_{f'})T_{2}/2)}{(E_{k}+E_{D}^{(k)}-E_{f'})T_{2}/2}
\end{equation}
and
\begin{equation}
\frac{sin((E_{k}+E_{D}^{(k)}-E_{f'})T_{2}/2)}{(E_{k}+E_{D}^{(k)}-E_{f'})T_{2}/2} \cdot \frac{sin((E_{k'}+E_{D}^{(k')}-E_{f'})T_{2}/2)}{(E_{k'}+E_{D}^{(k')}-E_{f'})T_{2}/2} \simeq sinc(\Delta_{k}/2)sinc((\Delta_{k}+\delta)T_{2} \slash 2)
\end{equation}
where $\Delta_{k}=E_{k}+E_{D}^{(k)}-E_{f'}$,

\begin{equation}
\delta=\Delta m^{2}_{kk'} \Big( \frac{\partial}{\partial m^{2}_{k}} \Big)_{m^{2}_{k}=m^{2}_{k'}} = \frac{\Delta m^{2}_{kk'}}{2E} \Big( \frac{v_{D,x}-v_{A,x}}{v_{k'}-v_{A,x}}  \Big) \simeq \frac{\Delta m^{2}_{kk'}}{2E} \Big( \frac{v_{D,x}-v_{A,x}}{1-v_{A,x}} \Big)
\end{equation}
$(52)$ and $(53)$ imply that there will be a loss of coherence between 2 neutrino mass modes, $k$ and $k'$, if
\begin{equation}
\abs{ \frac{\Delta m^{2}_{kk'}}{2E} \Big( \frac{v_{D,x}-v_{A,x}}{1-v_{A,x}} \Big) } > \frac{4\pi}{T_{2}}
\end{equation}
($\frac{4\pi}{T_{2}}$ is the width of $sin((\Delta_{k} T_{2})⁄2) \slash \Delta_{k}$ ).

   The phase difference between these 2 modes, which is responsible for probability oscillation, is given by
\begin{equation}
\phi_{k}-\phi_{k'} = \Delta m^{2}_{kk'} \cdot \Big( \frac{\partial \phi_{k}}{\partial m^{2}_{k}} \Big)_{m^{2}_{k}=0}
\end{equation}
where,
\begin{eqnarray}
  \frac{\partial \phi_{k}}{\partial m^{2}_{k}} = \lefteqn{ \Bigg[ -\frac{\partial E_{k}}{\partial m^{2}_{k}} (T_{1}+T_{prop}) + \frac{\partial p_{k}}{\partial m^{2}_{k}}L - \frac{\partial}{\partial m^{2}_{k}}(E_{k}+E_{D})\frac{T_{2}}{2} \Bigg] } \nonumber \\
&& + \Bigg[  \Big(  \frac{\partial}{\partial m^{2}_{k}} \frac{(E_{k}+E_{D}-E_{f'})}{2(v_{k}-v_{D,x})} \Big) \cdot \omega_{k} +\frac{(E_{k}+E_{D}-E_{f'})}{2(v_{k}-v_{D,x})} \cdot \frac{\partial \omega_{k}}{\partial m^{2}_{k}}  \Bigg]
\end{eqnarray}
The term in the first bracket in $(56)$ can be expressed as 
\begin{eqnarray}
\lefteqn{  -\frac{\partial E_{k}}{\partial m^{2}_{k}} (T_{1}+T_{prop}) + \frac{\partial p_{k}}{\partial m^{2}_{k}}L - \frac{\partial}{\partial m^{2}_{k}}(E_{k}+E_{D})\frac{T_{2}}{2} } \nonumber \\
&& = \frac{1}{2E_{k}} \cdot \frac{1}{\abs{v_{A,x}-v_{k}}} \Big[  v_{A,x}(T_{1}+T_{prop})-L+(v_{A,x}-v_{D,x}) \frac{T_{2}}{2} \Big]
\end{eqnarray}
From Table I, assuming $(47)$ and $(49)$, the first term in the second bracket of $(58)$ is given by 
\begin{eqnarray}
\lefteqn{ \Bigg(  \frac{\partial}{\partial m^{2}_{k}} \frac{(E_{k}+E_{D}-E_{f'})}{2(v_{k}-v_{D,x})} \Bigg) \cdot \omega_{k} } \nonumber \\
&\simeq&  \frac{1}{4E_{k} (v_{k}-v_{D,x})(v_{A,x}-v_{k})} \Bigg(  (v_{A,x}-v_{D,x}) - \frac{(E_{k}+E_{D}-E_{f'})}{(v_{k}-v_{D,x})} \Big[ \frac{1}{E_{k}} (1-v_{k}v_{A,x}) \nonumber \\
&& + \frac{1}{E_{D}}(1-v^{2}_{D,x}) \Big]  \Bigg) \cdot \Bigg( \frac{\Delta m^{2}_{Lk}}{2E^{2}} (T_{1}+T_{prop}) \Bigg) \nonumber \\
&\simeq& 0
\end{eqnarray}
(note that $E_{k}+E_{D}-E_{f'} \sim \mathcal{O}(1 \slash T_{2} )$), and second term in the second bracket is 
\begin{equation}
     \frac{(E_{k}+E_{D}-E_{f'})}{2(v_{k}-v_{D,x})}  \cdot \frac{\partial \omega_{k}}{\partial m^{2}_{k}} \simeq \frac{1}{2E^{2}} (T_{1}+T_{prop}) \cdot \frac{1}{(v_{k}-v_{D,x})T_{2}} \simeq 0
\end{equation}
which means that
\begin{eqnarray}
\lefteqn{ \phi_{k}-\phi_{k'} } \nonumber \\
&=& \frac{\Delta m^{2}_{kk'}}{2E_{k'}} \cdot \frac{1}{\abs{v_{A,x}-v_{k'}}} \Big[  v_{A,x}(T_{1}+T_{prop})-L+(v_{A,x}-v_{D,x}) \frac{T_{2}}{2}  \Big]  \nonumber \\
&=& \frac{\Delta m^{2}_{kk'}}{2E_{k'}} \cdot \frac{1}{\abs{v_{A,x}-v_{k'}}} \Big[  (v_{A,x}-v_{k'})(T_{1}+T_{prop})+(v_{A,x}-v_{D,x}) \frac{T_{2}}{2} +v_{k'}(T_{1}+T_{prop}) -L \Big]   \nonumber \\
&\simeq& \frac{\Delta m^{2}_{kk'}}{2E} \Big[   -L +\Big( \frac{v_{A,x}-v_{D,x}}{1-v_{A,x}} \Big) \frac{T_{2}}{2}  \Big] 
\end{eqnarray}
Using $(52)$ and $(60)$ in $(40)$, we have
\begin{eqnarray}
\nonumber \lefteqn{ P(\beta|\alpha,B;\vec{L}) } \\
&\simeq& \Big( \frac{2\pi T_{2}}{T_{1}} \Big)^{2} \sum_{k,k'} U^{\ast}_{\alpha k} U_{\beta k} U_{\alpha k'} U^{\ast}_{\beta k'}  \int D_{l_{\beta} \otimes C} \int d^{3}p_{D} \frac{1}{2E_{D}} \abs{ \tilde{\Psi}_{k,B} }^{2} \abs{ \Psi_{D} }^{2} e^{i \frac{\Delta m^{2}_{kk'}}{2E} \Big[   -L +\Big( \frac{v_{A,x}-v_{D,x}}{1-v_{A,x}} \Big) \frac{T_{2}}{2}  \Big] }  \nonumber \\ 
&& \cdot \int d \Delta_{2} sinc(\Delta_{2} T_{2}/2)sinc((\Delta_{2}+\delta) T_{2}/2)) \abs{iT^{D}_{C}}^{2} \delta(\Delta_{2}-\Delta_{k}) \delta^{3}(\vec{p}+\vec{p}_{D}-\vec{p}_{C}-\vec{p}_{l_{\beta}}) \nonumber \\
&=& \Big( \frac{T_{2}}{T_{1}} \Big)^{2} \sum_{k,k'} U^{\ast}_{\alpha k} U_{\beta k} U_{\alpha k'} U^{\ast}_{\beta k'}  \int d^{3}p_{D} \frac{1}{2E_{D}} \frac{2E \abs{1-v_{D,x}}}{(2\pi)^{3}} \abs{ \tilde{\Psi}_{k,B} }^{2} \abs{ \Psi_{D} }^{2} e^{i \frac{\Delta m^{2}_{kk'}}{2E} \Big[   -L +\Big( \frac{v_{A,x}-v_{D,x}}{1-v_{A,x}} \Big) \frac{T_{2}}{2}  \Big] }  \nonumber \\ 
&& \cdot \int d \Delta_{2} sinc(\Delta_{2} T_{2}/2)sinc((\Delta_{2}+\delta) T_{2}/2)) \sigma(E,\vec{p}_{D};l_{\beta},\Delta_{2}) 
\end{eqnarray}
where
\begin{eqnarray}
 \sigma(E,\vec{p}_{D};l_{\beta},\Delta_{2}) &\equiv& \frac{1}{4E_{D}E \abs{1-v_{D,x}}} \int D_{l_{\beta} \otimes C} (2\pi)^{4} \delta(E+E_{D}-E_{C}-E_{l_{\beta}}-\Delta_{2})  \nonumber \\
&& \cdot \delta^{3}(\vec{p}+\vec{p}_{D}-\vec{p}_{C}-\vec{p}_{l_{\beta}}) \abs{iT^{(D)}_{C}}^{2}
\end{eqnarray}
which is the standard definition \cite{Peskin} of the neutrino detection cross-section with the energy conservation modified by $\Delta_{2}$. This cross-section is assumed to be insensitive to neutrino masses ($(E,\vec{p})$ is the energy-momentum of the massless neutrino, $v_{k}=1$ in the Moller factor). Note that we have introduced $\delta^{3} (\vec{p}+\vec{p}_{D}-\vec{p}_{C}-\vec{p}_{l_{\beta}})$ into $(61)$ such that $\vec{p}_{D}$ is now an independent variable under $\int d^{3} p_{D}$, and that $\Delta_{2}$ is just a dummy variable related to $\Delta_{k}$ (defined just after $(52)$) only through $\delta(\Delta_{2}-\Delta_{k})$. The $T_{prop}$-dependence of $P(\beta|\alpha,B;\vec{L})$ is also dropped due to $(49)$.
   From $(26)$,
\begin{eqnarray}
  \mathcal{N}_{k} &\simeq& \Big[  \frac{16\pi^{3}}{\abs{(p_{B,x}+p_{k})E_{k}-p_{k}E_{A}}} \abs{\Psi_{A}(\vec{p}_{A})}^{2} \abs{iT^{(S)}_{B}}^{2} \int d \Delta_{1} sinc^{2}\Big( \frac{\Delta_{1}T_{1}}{2} \Big) \Big]^{1/2}  \nonumber \\
&=& \Big[  \frac{32\pi^{4}}{T_{1} \abs{(p_{B,x}+p_{k})E_{k}-p_{k}E_{A}}} \abs{\Psi_{A}(\vec{p}_{A})}^{2} \abs{iT^{(S)}_{B}}^{2}  \Big]^{1/2}
\end{eqnarray}
and substituting into $(25)$, we have
\begin{equation}
\abs{ \tilde{\Psi}_{k,B} }^{2} \simeq \frac{T_{1}}{32\pi^{4} E \abs{1-v_{A,x}}}
\end{equation}
Furthermore, if we once again assume that $ \sigma(E,\vec{p}_{D};l_{\beta},\Delta_{2})$ is smooth in $\Delta_{2}$ such that by the constraint of the $sinc$ function in $(61)$ we can set  $\sigma(E,\vec{p}_{D};l_{\beta},\Delta_{2}) \rightarrow  \sigma(E,\vec{p}_{D};l_{\beta},\Delta_{2}=0)\equiv  \sigma(E,\vec{p}_{D};l_{\beta})$ under the integral $\int d\Delta_{2}$ , the neutrino detection probability can be written as 
\begin{eqnarray}
\nonumber \lefteqn{ P(\beta|\alpha,B;\vec{L}) } \\
&\simeq&  \sum_{k,k'} U^{\ast}_{\alpha k} U_{\beta k} U_{\alpha k'} U^{\ast}_{\beta k'}   \int \frac{d^{3}p_{D}}{(2\pi)^{5}} \frac{ \abs{1-v_{D,x}}T_{2} }{ \abs{1-v_{A,x}}T_{1} }  \abs{ \Psi_{D} }^{2} e^{i \frac{\Delta m^{2}_{kk'}}{2E} \Big[   -L +\Big( \frac{v_{A,x}-v_{D,x}}{1-v_{A,x}} \Big) \frac{T_{2}}{2}  \Big] }  \nonumber \\ 
&& \cdot sinc \Big[ \frac{\Delta m^{2}_{kk'}}{4E}  \Big( \frac{v_{A,x}-v_{D,x}}{1-v_{A,x}} \Big) T_{2}  \Big] \cdot \sigma(E,\vec{p}_{D};l_{\beta})
\end{eqnarray}
One could see that the usual oscillation phase$ -\Delta m_{kk'}^{2} L \slash 2E$ is corrected by the term $\frac{\Delta m^{2}_{kk'}}{4E}  \times \Big( \frac{v_{A,x}-v_{D,x}}{1-v_{A,x}} \Big) T_{2} $, which is dependent on the relative velocities of the particles involved, and is constrained by the function $sinc \Big[ \frac{\Delta m^{2}_{kk'}}{4E}  \Big( \frac{v_{A,x}-v_{D,x}}{1-v_{A,x}} \Big) T_{2}  \Big]$. 
   Before we proceed further, it is prudent to check that the various factors entering the integral $(65)$ is consistent with the physical quantity that is being calculated, namely, the detection probability of a neutrino; having arrived at $(65)$ through numerous approximations and mathematical manipulations. This will also enable us to identify the quantities that will be more relevant to neutrino phenomenology.    

\subsection{Deconstructing $P(\beta|\alpha,B;\vec{L})$}

   Consider the case whereby the oscillatory correction term in $(65)$, is negligible; $\frac{\Delta m^{2}_{kk'}}{4E}  \Big( \frac{v_{A,x}-v_{D,x}}{1-v_{A,x}} \Big) T_{2}$ $\ll 1$, which implies (by consistency) $sinc \Big[ \frac{\Delta m^{2}_{kk'}}{4E}  \Big( \frac{v_{A,x}-v_{D,x}}{1-v_{A,x}} \Big) T_{2}  \Big] \simeq 1$. Thus the neutrino detection probability becomes
\begin{eqnarray}
\nonumber \lefteqn{ P(\beta|\alpha,B;\vec{L}) } \\
&\simeq&  \sum_{k,k'} U^{\ast}_{\alpha k} U_{\beta k} U_{\alpha k'} U^{\ast}_{\beta k'}   \int \frac{d^{3}p_{D}}{(2\pi)^{5}} \frac{ \abs{1-v_{D,x}}T_{2} }{ \abs{1-v_{A,x}}T_{1} }  \abs{ \Psi_{D} }^{2} e^{-i \frac{\Delta m^{2}_{kk'}L}{2E} }  \nonumber \\ 
&& \cdot \sigma(E,\vec{p}_{D};l_{\beta})
\end{eqnarray}
Using the standard definition of neutrino oscillation probability,
\begin{equation}
P(\nu_{\alpha} \rightarrow \nu_{\beta}) =  \sum_{k,k'} U^{\ast}_{\alpha k} U_{\beta k} U_{\alpha k'} U^{\ast}_{\beta k'}  e^{-i \frac{\Delta m^{2}_{kk'}L}{2E} }  
\end{equation}
and that of the detection cross-section
\begin{eqnarray}
P(l_{\beta} | E, \vec{p}_{D}) &=& (single-particle-flux) \times (cross-section) \times (time) \nonumber \\
&=& \frac{T_{2} \abs{1-v_{D,x}}}{V} \cdot \sigma(E,\vec{p}_{D};l_{\beta})
\end{eqnarray}
where $P(l_{\beta} | E, \vec{p}_{D})$ is the production probability of $l_{\beta}$ from the interaction between a neutrino of energy $E$ and a detection particle of momentum $\vec{p}_{D}$, over time $T_{2}$. $V$ is the volume of a single neutrino wave function.
   Substituting $(67)$ and $(68)$ back into $(66)$, we have
\begin{equation}
P(\beta|\alpha,B;\vec{L})  = \frac{V}{(2\pi)^{2}T_{1} \abs{1-v_{A,x}}}   \int \frac{d^{3}p_{D}}{(2\pi)^{3}}   \abs{ \Psi_{D} }^{2} P(\nu_{\alpha} \rightarrow \nu_{\beta}) P(l_{\beta} | E, \vec{p}_{D})  
\end{equation}

Apart from the pre-factor, $(69)$ has exactly the form of the average (over $\vec{p}_{D}$) production probability of $l_{\beta}$ from the capture of the neutrino, “$\nu_{\beta}$”, which has oscillated from its initial state, “$\nu_{\alpha}$”.

This pre-factor is an artifact of improper normalization of the incident neutrino wave packet in $(24)$. It is treated like a wave function in the one-dimensional phase space, instead of a three-dimensional one with zero transverse momentum. 

Since it has no dependence on the transverse neutrino momenta, a phase space normalization factor, $\sqrt{(2\pi)^{2} \slash \Sigma}$ , where $\Sigma$ is the transverse surface area should be included. This can be seen as the result of having to modify the neutrino wave packet in $(24)$ by the direct product:
\begin{equation}
\ket{\nu_{k}}_{new} = \ket{\nu_{k}} \otimes \ket{\vec{p}_{k,\perp}}
\end{equation}
such that:
\begin{eqnarray}
\braket{\nu_{k} | \nu_{k}}_{new} &=& \braket{ \nu_{k}  | \nu_{k}} \braket{\vec{p}_{k,\perp} | \vec{p}_{k,\perp}}  \nonumber \\
&=& \braket{\vec{p}_{k,\perp} | \vec{p}_{k,\perp}}
\end{eqnarray}
where $\ket{\vec{p}_{k,\perp}}$ is the transverse momentum state of the neutrino (with $\ket{\vec{p}_{k,\perp}}=0$). $\ket{\nu_{k}}$ is the neutrino state in $(24)$.

    $\ket{\vec{p}_{k,\perp}}$ is defined so that $\ket{\vec{p}_{k}}=\ket{p_{k}} \otimes \ket{\vec{p}_{k,\perp}}$ would satisfy the normalization condition $\braket{\vec{p}_{k} | \vec{p}_{k}'} = 2E_{k}(2\pi)^{3} \delta^{3}(\vec{p}_{k} - \vec{p}_{k}')$. This implies:
\begin{equation}
\braket{p_{k} | p_{k}'} = 2E_{k}(2\pi)^{3} \delta(p_{k} - p_{k}')
\end{equation}
\begin{equation}
\braket{\vec{p}_{k,\perp} | \vec{p}_{k,\perp}'} = \delta^{2}(\vec{p}_{k,\perp} - \vec{p}_{k,\perp}')
\end{equation}
Hence:
\begin{equation}
\braket{\vec{p}_{k,\perp} | \vec{p}_{k,\perp}} = \int \frac{d^{2}x_{\perp}}{(2\pi)^{2}} e^{i\vec{x}_{\perp} \cdot (\vec{p}_{k,\perp} - \vec{p}_{k,\perp})} = \int \frac{d^{2}x_{\perp}}{(2\pi)^{2}} = \frac{\Sigma}{(2\pi)^{2}}
\end{equation}
($\vec{x}_{\perp}$ is the transverse spatial components). In this last step, it is tacitly assumed that the spatial transverse area of the neutrino wave function is bounded ($=\Sigma$), this would suggest that there is a finite transverse momentum spread. This spread would have to be small in order that our approximation of $\vec{p}_{k,\perp} \simeq 0$ holds. This issue will be further discussed in the next section.

From $(71)$ and $(74)$, it follows that the normalized $\ket{\nu_{k}}_{new}$ is given by:
\begin{equation}
\ket{\nu_{k}}_{new} = \ket{\nu_{k}} \otimes \ket{\vec{p}_{k,\perp}} \cdot \sqrt{ \frac{(2\pi)^{2} }{\Sigma}}
\end{equation}
which implies that a corresponding factor, $\sqrt{(2\pi)^{2} \slash \Sigma}$, would have to be added to the amplitude, $\mathcal{A}(\beta|\alpha,B;C,\vec{L},T_{prop})$, and $(2\pi)^{2} \slash \Sigma$ to $P(\beta|\alpha,B;\vec{L})$.

Since the longitudinal width of the neutrino wave function is $\delta x_{\nu}=T_{1} \abs{1-v_{A,x}}$ (massless neutrino, according to $(38)$), we have:
\begin{equation}
\frac{\Sigma}{(2\pi)^{2}} = \frac{1}{(2\pi)^{2}} \frac{V}{\delta x_{\nu}} = \frac{1}{(2\pi)^{2}} \frac{V}{T_{1} \abs{1-v_{A,x}}}
\end{equation}
which exactly accounts for the pre-factor in $(69)$.

   Hence, we should have:
\begin{equation}
P(\beta|\alpha,B;\vec{L})  =  \int \frac{d^{3}p_{D}}{(2\pi)^{3}}   \abs{ \Psi_{D} }^{2} P(\nu_{\alpha} \rightarrow \nu_{\beta}) P(l_{\beta} | E, \vec{p}_{D})  
\end{equation}
A good feature of this treatment is that a normalized neutrino oscillation probability $(67)$, naturally emerge from the field theoretical calculation of the neutrino detection probability, if all the in-states of the scatterings are normalized; thus unitarity is explicit and not obscure by ad hoc normalization of probabilities, as mentioned in Section II. This completes our consistency check. 
  
   With the above considerations, $(65)$ becomes:
\begin{eqnarray}
\nonumber \lefteqn{ P(\beta|\alpha,E,\vec{p}_{A};\vec{L}) } \\
&\simeq&  \frac{1}{\Sigma \cdot T_{1} \abs{1-v_{A,x}}} \int d^{3} p_{D} F(\vec{p}_{D})  \abs{1-v_{D,x}}T_{2} \sigma(E,\vec{p}_{D};l_{\beta}) \nonumber \\
 && \cdot  \sum_{k,k'} U^{\ast}_{\alpha k} U_{\beta k} U_{\alpha k'} U^{\ast}_{\beta k'}  e^{i \frac{\Delta m^{2}_{kk'}}{2E} \Big[   -L +\Big( \frac{v_{A,x}-v_{D,x}}{1-v_{A,x}} \Big) \frac{T_{2}}{2}  \Big] }  \cdot sinc \Big[ \frac{\Delta m^{2}_{kk'}}{4E}  \Big( \frac{v_{A,x}-v_{D,x}}{1-v_{A,x}} \Big) T_{2}  \Big]
\end{eqnarray}
where the change of notation, $P(\beta | \alpha,B;\vec{L}) \rightarrow P(\beta | \alpha,E,\vec{p}_{A};\vec{L})$, reflects the independent kinematic variables in a more convenient way. $F(\vec{p}_{D})$ is the normalized momentum distribution function of $D$ ($\int d^3 p_{D} F(\vec{p}_{D})=1$), which realistically, will be a classical distribution rather than due to its wave function. The only ill-defined quantity in $(78)$ is $\Sigma$, which is the effective transverse area. This will be explained in the next section.

\subsection{Neutrino count rate}

From $(68)$ and $(76)$, $\Sigma$ can be taken to be the transverse surface area, within which the transversal displacement of $D$ (relative to the neutrino) is bounded, during the interaction time $T_{2}$, when the wave functions of the neutrino and $D$ are overlapping. Since this is not a measured quantity, a relevant quantity should be derived from $(78)$, whereby $\Sigma$ drops out. A good candidate is a probability rate defined as follows
\begin{equation}
\frac{dP(l_{\alpha},l_{\beta} | \vec{L})}{dt} \equiv \int d^3 p_{D} F(\vec{p}_{D}) \int dE \frac{d[\Gamma_{A}(\vec{p}_{A} \cdot T_{1})]}{dE d\Omega} \Delta \Omega \cdot \frac{dP(\beta|\alpha,E,\vec{p}_{A};\vec{L})}{dt}
\end{equation}
where,
\begin{equation}
\frac{dP(\beta|\alpha,E,\vec{p}_{A};\vec{L})}{dt} \equiv \frac{P(\beta|\alpha,E,\vec{p}_{A};\vec{L})}{T_{2}}
\end{equation}
$\Gamma_{A}(\vec{p}_{A}) \cdot T_{1}$ is the decay fraction of the source particle (with momentum $\vec{p}_{A}$) in time $T_{1}$.  $d\Gamma_{A}(\vec{p}_{A}) \slash dEd\Omega$ is the differential decay rate producing a neutrino of energy $E$, travelling within the solid angle $\Delta \Omega$, and a charged lepton, $l_{\alpha}$. $F(\vec{p}_{D})$ is the momentum distribution of $A$. $\Delta \Omega$ is the solid angle projected from the source to $D$, at which it covers the effective area $\Sigma$, such that
\begin{equation}
\Sigma=\Delta \Omega \times L^{2}
\end{equation}
With $(80)$ and $(81)$, $(79)$ can be expressed as
\begin{eqnarray}
\lefteqn{ \frac{dP(l_{\alpha},l_{\beta} | \vec{L})}{dt} \equiv \int d^3 p_{D} F(\vec{p}_{D}) \int dE \frac{d[\Gamma_{A}(\vec{p}_{A})]}{dE d\Omega}  \frac{1}{L^{2} \abs{1-v_{A,x}}} \int d^{3} p_{D} F(\vec{p}_{D}) } \nonumber \\
&&  \cdot \abs{1-v_{D,x}} \sigma(E,\vec{p}_{D};l_{\beta}) \nonumber \\
 && \cdot  \sum_{k,k'} U^{\ast}_{\alpha k} U_{\beta k} U_{\alpha k'} U^{\ast}_{\beta k'}  e^{i \frac{\Delta m^{2}_{kk'}}{2E} \Big[   -L +\Big( \frac{v_{A,x}-v_{D,x}}{1-v_{A,x}} \Big) \frac{T_{2}}{2}  \Big] }  \cdot sinc \Big[ \frac{\Delta m^{2}_{kk'}}{4E}  \Big( \frac{v_{A,x}-v_{D,x}}{1-v_{A,x}} \Big) T_{2}  \Big]
\end{eqnarray}
$\Sigma$ drops out of $dP(l_{\alpha},l_{\beta} | \vec{L}) \slash dt$. 

   From the definition $(79)$, it is apparent that the neutrino count rate ($dN_{l_{\beta}} \slash dt$) is related to $dP(l_{\alpha},l_{\beta} | \vec{L}) \slash dt$ by the relation 
\begin{equation}
\frac{dN_{l_{\beta}}}{dt} = \int_{\mathcal{V}_{S} \otimes \mathcal{V}_{D}} d^{3}x_{S} d^{3}x_{D} \rho_{S} (\vec{x}_{S}) \rho_{D} (\vec{x}_{D}) \frac{dP(l_{\alpha},l_{\beta} | \vec{L})}{dt}
\end{equation}
where $\mathcal{V}_{S} \slash \mathcal{V}_{D}$ is the volume of the source/detector, and $\rho_{S} (\vec{x}_{S}) \slash \rho_{D} (\vec{x}_{D})$ are their respective particle number densities. In term of source and detector coordinates, the propagation vector is $\vec{L}=\vec{x}_{D}-\vec{x}_{S}$.

\section{Discussion and conclusion}

   From $(82)$, it can be seen that the maximum allowable correction to the standard oscillation phase, $\abs{ \frac{\Delta m^{2}_{kk'}}{4E} \Big( \frac{v_{A,x}-v_{D,x}}{1-v_{A,x}} \Big) T_{2} }$, is limited by the width of the constraint function, $sinc \Big[\frac{\Delta m^{2}_{kk'}}{4E}  \times \Big( \frac{v_{A,x}-v_{D,x}}{1-v_{A,x}} \Big) T_{2} \Big]$, to be up to $\mathcal{O}(\pi)$. In fact, for $\abs{ \frac{\Delta m^{2}_{kk'}}{4E} \Big( \frac{v_{A,x}-v_{D,x}}{1-v_{A,x}} \Big) T_{2} } \leq \pi $, the expression
\begin{eqnarray}
\nonumber \lefteqn{ P(\nu_{\alpha} \rightarrow \nu_{\beta}) } \\
&\equiv&  \sum_{k,k'} U^{\ast}_{\alpha k} U_{\beta k} U_{\alpha k'} U^{\ast}_{\beta k'}  e^{i \frac{\Delta m^{2}_{kk'}}{2E} \Big[   -L +\Big( \frac{v_{A,x}-v_{D,x}}{1-v_{A,x}} \Big) \frac{T_{2}}{2}  \Big] }  \cdot  sinc \Big[ \frac{\Delta m^{2}_{kk'}}{4E}  \Big( \frac{v_{A,x}-v_{D,x}}{1-v_{A,x}} \Big) T_{2}  \Big] 
\end{eqnarray}
which is a modification of the standard expression $(67)$ as the result of $(78)$, can still be interpreted as the neutrino flavor oscillation probability (non-negative and normalized). 

   Under conditions where oscillation is observable, that is $\Delta m^{2}_{kk'}L \slash 4E \simeq \mathcal{O}(2\pi)$, this correction could still be significant. Recall that due to the locality of the neutrino absorption interaction, $(48)$ holds, and this leads to the upper bound of $\abs{v_{k}-v_{D,x}}T_{2} \simeq \abs{v_{A,x}-v_{k}}T_{1}$, which corresponds to a maximum phase correction of $\frac{\Delta m^{2}_{kk'} T_ {1}}{4E}  \Big( \frac{v_{A,x}-v_{D,x}}{1-v_{A,x}} \Big)$. This is equivalent to a shift in the neutrino propagation length $L$; $L \rightarrow L-\Big(  \frac{v_{A,x}-v_{D,x}}{1-v_{A,x}}  \Big) \frac{T_{1}}{2}$. In situations where the detector particles are of very low velocities, this shift reduces to $\simeq \frac{v_{A,x} T_{1}}{2}$, which is equal to half the distance travelled by the source particle in the decoherence time between its decayed and undecayed states. 

   This could be significant in the case of a beam source (for example a pion beam) in a short baseline experiment, where this distance, if assumed to be of the order of magnitude of the source decay tunnel, would not be negligible, compared to the subsequent distance travelled by the neutrino to the detector. An example is the MiniBooNE experiment \cite{Stancu}, where tunnel length is $50m$, while the distance from the end of the tunnel to the detector $\sim 450m$.

   On the other hand, if the decoherence time is fast, such that the distance travelled by the source particle is negligible compared to the propagation length of the neutrino, the phase correction term could be dropped, with $sinc \Big[ \frac{\Delta m^{2}_{kk'}}{4E}  \Big( \frac{v_{A,x}-v_{D,x}}{1-v_{A,x}} \Big) T_{2}  \Big] \rightarrow 0$. The modified oscillation formula $(84)$ thus reduces to the standard expression. 

   This could result from the electromagnetic interaction between the charged source particles or with the magnetic field that focused them into a beam. The corresponding decoherence time ($T_{1}$) would be the time taken, for the difference between the perturbation of the environment by a source particle and its daughter charged lepton, to build up significantly. The distance travelled by the source in this time should be compared with the dimensions of the decay tunnel, and the smaller of these will determine the significance of the corrections to the standard oscillation formula discussed here.      

\appendix*
\section{Simplifying $\mathcal{A}(\beta|\alpha,B;C,\vec{L},T_{prop})$: some approximations}

To further simplify the expression for the neutrino detection amplitude, $\mathcal{A}(\beta|\alpha,B;C,\vec{L},T_{prop})$, so as to facilitate the evaluation of the detection probability
\begin{equation}
P(\beta|\alpha,B;\vec{L},T_{prop})=\int D_{l_{\beta} \otimes C} \abs{\mathcal{A}(\beta|\alpha,B;C,\vec{L},T_{prop})}^{2}
\end{equation}
($\int D_{l_{\beta} \otimes C}$  sums the final states of $(3)$), some assumptions have to be made. The first is regarding the wave packet size of the source and detector particle, $A$ and $D$. We assume that the spatial dimensions of the wave packets of these particles to be much small than the decoherence times ($T_{1}$ and $T_{2}$. They are also assumed to be of the same orders of magnitude).

For example consider $A$ in the source under collisional monitoring. Thus $T_{1}$ could be comparable to the mean free path (or much large, if $A$ is not traveling at relativistic speed) or multiples of it (if multiple collisions is required for the decoherence of the state in $(11)$ to occur). This assumption means that $\delta x_{A} \ll T_{1}$ ($\delta x_{A}$ is the spatial size of the wave packet of $A$). This is reasonable purely from the definition of a mean free path. In momentum space, this means 
\begin{equation}
\delta p_{A} \gg \frac{1}{T_{1}}
\end{equation}
which says that the momentum wave function, $\Psi_{A} (\vec{p}_{A} )$, is much broader than the function $sinc(\Delta_{1} T_{1} \slash 2)$ in $(31)$. This is assumed to be also true for $\Psi_{D} (\vec{p}_{D} )$.

Thus we shall set $\Delta_{1}=0$ for all functions except $sinc(\Delta_{1} T_{1} \slash 2)$, and for those functions with components (for example phases) that could accumulate with space and time ($T_{1}$, $T_{prop}$, $T_{2}$ and $\vec{L}$, which are all potentially macroscopic quantities), we approximate linearly in $\Delta_{1}$. 

   Implementing these approximations on $(31)$, we obtain
\begin{eqnarray}
\nonumber \lefteqn{ \mathcal{A}(\beta|\alpha,B;C,\vec{L},T_{prop}) } \\
&\simeq&T_{2} \sum_{k} U^{\ast}_{\alpha k} U_{\beta k} \sum_{s}  \Bigg[ \frac{1}{\sqrt{2E_{D}}} \tilde{\Psi}_{k,B}(\Delta_{1}) \Psi_{D}(\vec{p}_{D}) e^{-i\vec{p}_{f'} \cdot \vec{L}} e^{-iE_{f'}T_{2}/2}  \nonumber \\ 
&& \cdot iT^{(D)}_{C}\big((\vec{p}_{l_{\beta}},s_{l_{\beta}});(\vec{p}_{k}(\{\vec{p}_{B}\},\Delta_{1}),s), \vec{p}_{D} \big) \Bigg]_{\vec{p}_{D}=\vec{p}_{f'}-\vec{p}_{k}, \Delta_{1}=0} \nonumber \\
&& \int d\Delta_{1} \Bigg[ e^{ip_{k}L} e^{-i(\Delta_{1}T_{1}/2+E_{k}(\vec{p}_{k}(\{\vec{p}_{B}\},\Delta_{1}))(T_{1}+T_{prop}))} e^{-i(E_{k}(\vec{p}_{k}(\{\vec{p}_{B}\},\Delta_{1}))+E_{D})T_{2}/2} \nonumber \\
&& \cdot sinc\Big(\frac{\Delta_{1} T_{1}}{2}\Big)sinc\big( ( E_{k}(\vec{p}_{k}(\{\vec{p}_{B}\},\Delta_{1}))+E_{D}-E_{f'})T_{2}/2 \big) \Bigg]_{\vec{p}_{D}=\vec{p}_{f'}-\vec{p}_{k}(\{\vec{p}_{B}\}, \Delta_{1})}
\end{eqnarray}
where we have factored the terms evaluated at $\Delta_{1}=0$, out of the integral $\int d\Delta_{1}$. For the remaining term under the integral, we shall expand all energy-momenta (that are dependent on $\Delta_{1}=0$ through $\vec{p}_{k}(\{\vec{p}_{B}\},\Delta_{1})$ to the first order in $\Delta_{1}$. This results in 
\begin{eqnarray}
\nonumber \lefteqn{ \mathcal{A}(\beta|\alpha,B;C,\vec{L},T_{prop}) } \\
&\simeq&T_{2} \sum_{k} U^{\ast}_{\alpha k} U_{\beta k} \sum_{s}  \Bigg[ \frac{1}{\sqrt{2E_{D}}} \tilde{\Psi}_{k,B}(\Delta_{1}) \Psi_{D}(\vec{p}_{D}) e^{-i\vec{p}_{f'} \cdot \vec{L}} e^{iE_{f'}T_{2}/2}  \nonumber \\ 
&& \cdot iT^{(D)}_{C}\big((\vec{p}_{l_{\beta}},s_{l_{\beta}});(\vec{p}_{k}(\{\vec{p}_{B}\},\Delta_{1}),s), \vec{p}_{D} \big) \Bigg]_{\vec{p}_{D}=\vec{p}_{f'}-\vec{p}_{k}} \nonumber \\
&& \cdot \int d\Delta_{1} e^{i(G+F\Delta_{1})} sinc(f\Delta_{1})sinc(b\Delta_{1}+c)
\end{eqnarray}
Henceforth, for notational expedience, we shall assume implicit dependence of energy-momenta on the set $\{\vec{p}_{B} \}$ and that $\Delta_{1}=0$, unless otherwise stated (for example $\vec{p}_{k}≡\vec{p}_{k} (\{\vec{p}_{B} \}$, $\Delta_{1}=0)$). The new symbols introduced in (A.4) are
\begin{eqnarray}
G &=& \Big[-E_{k}(T_{1}+T_{prop})+p_{k}L-(E_{k}+E_{D})T_{2}/2) \Big]_{\vec{p}_{D}=\vec{p}_{f'}-\vec{p}_{k}}, \nonumber \\
F &=& \Big[\frac{\partial}{\partial \Delta_{1}} \big( -E_{k}(T_{1}+T_{prop})+p_{k}L-(E_{k}+E_{D})T_{2}/2) \big) \Big]_{\vec{p}_{D}=\vec{p}_{f'}-\vec{p}_{k}}-\frac{T_{1}}{2}, \nonumber \\
f &=& \frac{T_{1}}{2}, c=\frac{T_{2}}{2}\big[ E_{k}+E_{D}-E_{f'} \big]_{\vec{p}_{D}=\vec{p}_{f'}-\vec{p}_{k}},
b = \frac{T_{2}}{2} \Big[ \frac{\partial}{\partial \Delta_{1}}(E_{k}+E_{D} ) \Big]_{\vec{p}_{D}=\vec{p}_{f'}-\vec{p}_{k}} 
\end{eqnarray}
which corresponds to the coefficients of the linear expansions in $\Delta_{1}$. 

   The integral in (A.4) can be solved analytically (using the convolution theorem of Fourier transforms) and is given by
\begin{eqnarray}
I
&\equiv& \int d\Delta_{1} e^{i(G+F\Delta_{1})} sinc(f\Delta_{1})sinc(b\Delta_{1}+c) \nonumber \\
&=& \frac{\pi}{2\abs{b} \abs{f}} e^{-iF \frac{c}{b}} \int dk e^{ik \frac{c}{b}} \cdot rect \Bigg( \frac{k}{\abs{f}} \Bigg) rect \Bigg( \frac{k-F}{\abs{b}} \Bigg)
\end{eqnarray}
The function, $rect(k)$, is defined by
\begin{equation}
rect(k)=
\begin{cases}
1 ; -1 \le k \le 1\\
0 ; elsewhere
\end{cases}
\end{equation}
For $I \neq 0$, (A.7) can be written as
\begin{eqnarray}
I
&=& \frac{\pi}{2\abs{b} \abs{f}} e^{-iF \frac{c}{b}} \int_{k_{1}}^{k_{2}} dk e^{ik \frac{c}{b}} \nonumber \\
&=& \frac{\pi sgn(b)}{c \abs{f}} e^{i\big( -F \frac{c}{b}+\frac{c}{2b}(k_{1}+k_{2}) \big)} sin \Big[ \frac{c}{2b}(k_{2}-k_{1}) \Big]
\end{eqnarray}
The integration limits, $k_{1}$ and $k_{2}$, in (A.8) can be classified into the following cases (where $I \neq 0$):
\begin{enumerate}
\item For $\abs{b} > \abs{f}$, there are 3 cases:
  \begin{enumerate}
   \item[a.)] $-\abs{f}<F-\abs{b}<\abs{f}$, where $k_{1}=F-\abs{b}$ and $k_{2}=\abs{f}$ (see Fig. 5) 
   \item[b.)] $\abs{f}-2\abs{b}<F-\abs{b}<-\abs{f}$, where $k_{1}=-\abs{f}$ and $k_{2}=\abs{f}$ (see Fig. 6)
   \item[c.)] $-\abs{f}<F+\abs{b}<\abs{f}$, where $k_{1}=-\abs{f}$ and $k_{2}=F+\abs{b}$ (see Fig. 7)  
  \end{enumerate}
\item For $\abs{b} < \abs{f}$, there are 3 cases:
  \begin{enumerate}
   \item[d.)] $\abs{f}-2\abs{b}<F-\abs{b}<\abs{f}$, where $k_{1}=F-\abs{b}$ and $k_{2}=\abs{f}$ (see Fig. 8) 
   \item[e.)] $-\abs{f}<F-\abs{b}<\abs{f}$, where $k_{1}=F-\abs{b}$ and $k_{2}=F+\abs{b}$ (see Fig. 9)
   \item[f.)] $-\abs{f}-2\abs{b}<F-\abs{b}<-\abs{f}$, where $k_{1}=-\abs{f}$ and $k_{2}=F+\abs{b}$ (see Fig. 10)  
  \end{enumerate}
\end{enumerate}

\begin{figure}[h]
\centering
\includegraphics[scale=0.6]{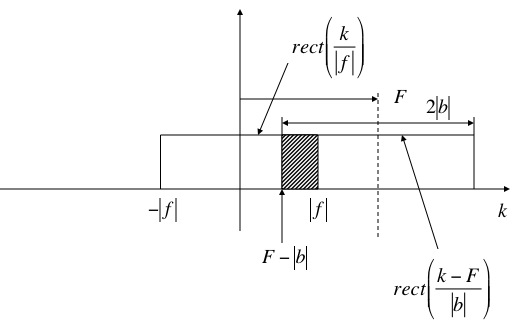}
\caption{\label{fig:fig5} a.) The dashed line is the center-line of $rect \Big( \frac{k-F}{\abs{b}} \Big)$. The shaded region represents the domain of integration in (A.8); $k_{1}=F-\abs{b}$, $k_{2}=\abs{f}$, for $-\abs{f}<F-\abs{b}<\abs{f}$, $\abs{b} > \abs{f}$.}
\end{figure}

\begin{figure}[h]
\centering
\includegraphics[scale=0.6]{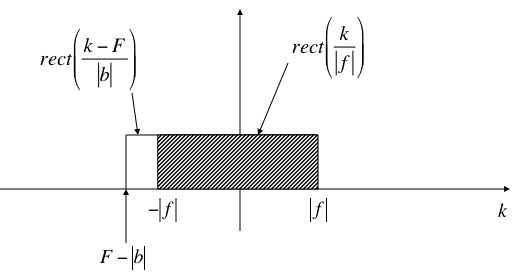}
\caption{\label{fig:fig6} b.) $k_{1}=-\abs{f}$, $k_{2}=\abs{f}$, for $\abs{f}-2\abs{b}<F-\abs{b}<-\abs{f}$, $\abs{b} > \abs{f}$.}
\end{figure}

\begin{figure}[h]
\centering
\includegraphics[scale=0.6]{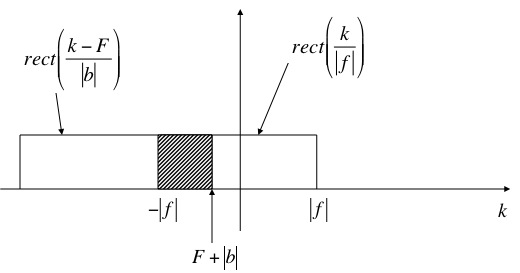}
\caption{\label{fig:fig7} c.) $k_{1}=-\abs{f}$, $k_{2}=F+\abs{b}$, for $-\abs{f}<F+\abs{b}<\abs{f}$, $\abs{b} > \abs{f}$.}
\end{figure}

\begin{figure}[h]
\centering
\includegraphics[scale=0.6]{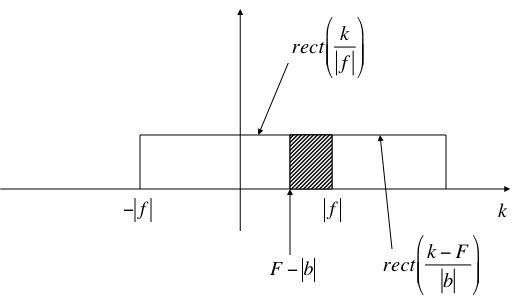}
\caption{\label{fig:fig8} d.) $k_{1}=F-\abs{b}$, $k_{2}=\abs{f}$, for $\abs{f}-2\abs{b}<F-\abs{b}<\abs{f}$, $\abs{b} < \abs{f}$.}
\end{figure}

\begin{figure}[h]
\centering
\includegraphics[scale=0.6]{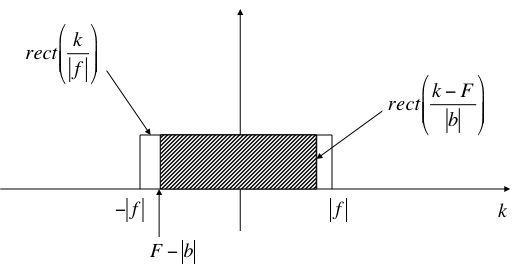}
\caption{\label{fig:fig9} e.) $k_{1}=F-\abs{b}$, $k_{2}=F+\abs{b}$, for $-\abs{f}<F-\abs{b}<\abs{f}$, $\abs{b} < \abs{f}$.}
\end{figure}

\begin{figure}[h]
\centering
\includegraphics[scale=0.6]{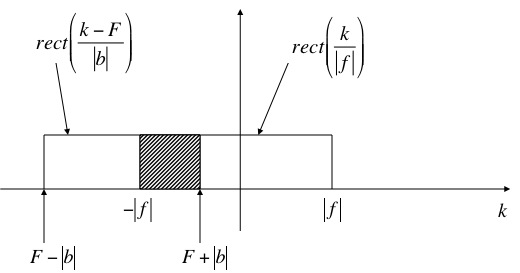}
\caption{\label{fig:fig10} f.) $k_{1}=-\abs{f}$, $k_{2}=F+\abs{b}$, for $-\abs{f}-2\abs{b}<F-\abs{b}<-\abs{f}$, $\abs{b} < \abs{f}$.}
\end{figure}

Summarizing the result for $I$,
\begin{equation}
I=
\begin{cases}
\frac{\pi sgn(b)}{c \abs{f}} e^{i\big( -F \frac{c}{b}+\frac{c}{2b}(k_{1}+k_{2}) \big)} sin \Big[ \frac{c}{2b}(k_{2}-k_{1}) \Big] ;$ a.), b.), c.), d.), e.) and f.)$ \\
0 ; \abs{F}>\abs{f}+\abs{b}
\end{cases}
\end{equation}
where $I=0$ when the $rect( )$ functions in (A.6) are not overlapping. 

   The coefficients involving the derivative, $(\partial \slash \partial \Delta_{1} )_{\Delta_{1}=0}$, in (A.5) ($F$ and $b$), can be expressed in terms of particle velocities
\begin{eqnarray}
F &=& \frac{-v_{k}}{v_{A,x}-v_{k}}(T_{1}+T_{prop})+\frac{L}{v_{A,x}-v_{k}}-\Big( \frac{v_{k}-v_{D,x}}{v_{A,x}-v_{k}} \Big)\frac{T_{2}}{2}- \frac{T_{1}}{2} \nonumber \\
b &=&\Big( \frac{v_{k}-v_{D,x}}{v_{A,x}-v_{k}} \Big)\frac{T_{2}}{2} 
\end{eqnarray}

 This results from the fact that the derivatives $(\partial \slash \partial \Delta_{1} )_{\Delta_{1}=0}$ (keeping $\{\vec{p}_{B} \}$ constant) of particle energy-momenta correspond to velocity ratios, which can be derived from the energy-momentum conservation equations $(15)$. For example, to find $(\partial p_{k} \slash \partial \Delta_{1} )_{\Delta_{1}=0}$, the first equation in $(15)$ could be used to eliminate $\vec{p}_{A}$ in the second equation, followed by the variation $\delta_{\Delta_{1}}$, 
\begin{eqnarray}
\delta_{\Delta_{1}} \sqrt{(p_{k}+p_{B,x})^{2}+p^{2}_{B,\perp}+m^{2}_{A}} &=& \delta_{\Delta_{1}} \Big( \Delta_{1}+\sqrt{p^{2}_{k}+m^{2}_{k}} +E_{B} \Big) \nonumber \\
 \frac{p_{k}+p_{B,x}}{ \sqrt{(p_{k}+p_{B,x})^{2}+p^{2}_{B,\perp}+m^{2}_{A}}} \delta_{\Delta_{1}} p_{k} &=& \delta \Delta_{1} + \frac{p_{k}}{\sqrt{p^{2}_{k}+m^{2}_{k}}} \delta_{\Delta_{1}} p_{k}
\end{eqnarray}
($p_{B,\perp}$  is the momentum component perpendicular to $p_{B,x}$). From this we get
\begin{equation}
\Big(\frac{\partial p_{k}}{\partial \Delta_{1}} \Big)_{\Delta_{1}=0}=\frac{1}{v_{A,x}-v_{k}}
\end{equation}
Similarly,
\begin{equation}
\Big(\frac{\partial E_{k}}{\partial \Delta_{1}} \Big)_{\Delta_{1}=0}=\frac{v_{k}}{v_{A,x}-v_{k}}, \Big(\frac{\partial E_{D}}{ \partial \Delta_{1}} \Big)_{\Delta_{1}=0}=\frac{-v_{D,x}}{v_{A,x}-v_{k}} 
\end{equation}

Using (A.5) and (A.10) on (A.9), we can rewrite the condition for zero neutrino detection amplitude as
\begin{eqnarray}
\mathcal{A}(\beta|\alpha,B;C,\vec{L},T_{prop})=0; \lefteqn{v_{k}(T_{1}+T_{prop}) < L-T_{2} \abs{v_{k}-v_{D,x}},} \nonumber \\
& v_{k}(T_{1}+T_{prop}) > L+T_{1} \abs{v_{A,x}-v_{k}}
\end{eqnarray}
The physical meaning of this condition is discussed in Section III.C.

\begin{acknowledgments}
This work is supported by the National University of Singapore academic research grant: no. WBS: R-144-000-178-112.
\end{acknowledgments}

\end{document}